 \documentclass[preprint2]{aastex}
\usepackage{graphics,epsfig,longtable,lscape,txfonts,color}
\usepackage[english]{babel}

\newcommand{\myemail}{hstiele@mx.nthu.edu.tw}

\newcommand{\mr}{\mathrm}
\newcommand{\nh}{\hbox{$N_{\mr H}$}}
\newcommand{\hcm}[1]{$\times 10^{#1}$ cm$^{-2}$}

\newcommand{\ergcm}[1]{$\times10^{#1}$~\hbox{erg~cm$^{-2}$~s$^{-1}$}}

\def\eg{e.\,g.}                                      

\def\xmm{\textit{XMM-Newton}}
\def\swift{\textit{Swift}}
\def\nus{\textit{NuSTAR}}
\def\maxi{MAXI~J1543-564}
\def\gx339{GX\,339-4}
\def\h1743{H\,1743-322}

\def\max15{MAXI\,J1535--571}

\shorttitle{2017 observations of \max15}
\shortauthors{Stiele, Kong}

\begin{document}


\title{A spectral and timing study of \max15, based on \swift/XRT, \xmm\ and NICER observations obtained in fall 2017}

\author{H.\ Stiele}
\affil{Institute of Astronomy, National Tsing Hua University, No.~101 Sect.~2 Kuang-Fu Road,  30013, Hsinchu, Taiwan}
\email{\myemail}
\and

\author{A.\ K.\ H.\ Kong}
\affil{Institute of Astronomy, National Tsing Hua University, No.~101 Sect.~2 Kuang-Fu Road,  30013, Hsinchu, Taiwan}





\begin{abstract}
We present a spectral-timing analysis of observations taken in fall 2017 of the newly detected X-ray transient \max15. We included 38 \swift/XRT window timing mode observations, three \xmm\ observations and 31 NICER observations in our study. We computed the fundamental diagrams commonly used to study black hole transients, and fitted power density and energy spectra to study the evolution of spectral and timing parameters. The observed properties are consistent with a bright black hole X-ray binary ($F_{\mr{0.6-10 keV}}^{\mr{max}}=3.71\pm0.02$\ergcm{-7}) that evolves from the low-hard-state to the high-soft state and back to the low-hard-state. In some observations the power density spectra showed type-C quasi-periodic oscillations, giving additional evidence that \max15\ is in a hard state during these observations. The duration of the soft state with less than ten days is unusually short and observations taken in spring 2018 show that \max15\ entered a second (and longer) soft state.
\end{abstract}


\section{Introduction}
Most low-mass black hole X-ray binaries are transient sources that evolve through different states during an outburst \citep{2006csxs.book..157M,2010LNP...794...53B}. The evolution during their outbursts can be studied using hardness intensity diagram \citep[HID;][]{2001ApJS..132..377H,2005A&A...440..207B,2005Ap&SS.300..107H,2006MNRAS.370..837G,2006csxs.book..157M,2009MNRAS.396.1370F,2010LNP...794...53B,2011BASI...39..409B}, hardness root-mean square (rms) diagram \citep[HRD;][]{2005A&A...440..207B} and rms intensity diagram \citep[RID;][]{2011MNRAS.410..679M}. In the low-hard state (LHS), rms of several tens of per cent is observed and the emission is dominated by thermal Comptonization in a hot, geometrically thick, optically thin plasma located in the vicinity of the black hole, where softer seed photons coming from an accretion disk are up-Comptonized \citep[see][for reviews]{2007A&ARv..15....1D,2010LNP...794...17G}. In the high-soft state (HSS), the variability is much lower \citep[fractional rms $\sim$1 per cent, e.g.][]{2005A&A...440..207B} and the spectrum is clearly dominated by an optically thick, geometrically thin accretion disk \citep{1973A&A....24..337S}.

A detailed overview of the properties of different types of QPOs and their relation to different outburst states can be found in \citet{2014SSRv..183...43B}. Here we just give a short summary: In the LHS and hard intermediate state (HIMS), type-C QPOs can be present \citep[][and references therein]{1999ApJ...514..939W,2011MNRAS.418.2292M}. These oscillations are observed in a large number of sources \citep{2006csxs.book..157M,2010LNP...794...53B}, have centroid frequencies ranging from 0.01 to 30 Hz, and their quality factor ($Q=\nu_0/(2\Delta)$, where $\nu_0$ is the centroid frequency, and $\Delta$ is the half width at half maximum) is $\ga10$ \citep[see \eg\ ][]{2005ApJ...629..403C,2010ApJ...714.1065R}. Often these oscillations appear with one or two overtones and at times with a sub-harmonic. They are always associated to a band limited noise and its frequency is anti-correlated with the total broad-band fractional rms variability. The soft intermediate state (SIMS) is defined by the presence of weaker power-low noise and of type-B QPOs. These oscillations have centroid frequencies of 0.8 -- 6.4 Hz, $Q>6$, have a 5 -- 10\% fractional rms and appear often together with an overtone and a sub-harmonic. Type-A QPOs have centroid frequencies of 6.5 -- 8 Hz, are broad ($Q\sim1-3$) and weak (fractional rms $<5$\%).  The three types of QPOs are well separated as a function of the total integrated fractional rms in the power density spectrum.

In this paper, we present a comprehensive study of the spectral and temporal variability properties of \max15\ observed during its 2017 outburst. On September 2, 2017 MAXI/GSC \citep{2017ATel10699....1N} and \swift/BAT \citep{2017ATel10700....1K} detected a bright uncatalogued  hard X-ray transient located near the Galactic plane. This source, \max15, has been classified as black hole X-ray binary candidate based on its behaviour observed in monitoring X-ray and radio observations \citep{2017ATel10708....1N,2017ATel10711....1R}.  State transitions and the detection of quasi-periodic oscillations (QPOs) have been reported \citep{2017ATel10731....1K,2017ATel10729....1N,2017ATel10734....1M}. \nus\ spectra obtained five days after the detection of \max15\ reveal the presence of a strong reflection component \citep{2018ApJ...852L..34X}.  

\section[]{Observation and data analysis}
\label{Sec:obs}
\subsection{Neil Gehrels Swift Observatory}
\label{Sec:obs_sw}
\max15\ was detected in \swift/BAT and MAXI/GSC monitoring observations on 2017 September 2nd \citep{2017GCN.21788....1M,2017ATel10699....1N}. 
We analysed all \swift/XRT \citep{2005SSRv..120..165B} monitoring data of \max15\ obtained in window timing mode between September 2nd and October 24th, excluding one observation which has an exposure shorter than 100~s. We extracted energy spectra of each observation using the online data analysis tools provided by the Leicester \swift\ data centre\footnote{http://www.swift.ac.uk/user\_objects/}, including single pixel events only. Background spectra are produced from the entire window, excluding a 120-pixel wide box centred on the source \citep{2009MNRAS.397.1177E}.  

In addition, we extracted power density spectra (PDS) in the 0.3 -- 10 keV energy band, following the procedure outlined in \citet{2006MNRAS.367.1113B}. We subtracted the contribution due to Poissonian noise \citep{1995ApJ...449..930Z}, normalised the PDS according to \citet{1983ApJ...272..256L} and converted to square fractional rms \citep{1990A&A...227L..33B}. The contribution due to Poissonian noise is determined by fitting the flat tail of the PDS at the high-frequency end with a constant. Determining the value of the Poissonian noise contribution that way, allows to take into account deviation from the expected value of 2, that are caused by pile-up effects in the \swift/XRT data \citep{2013ApJ...766...89K}. The PDS were fitted with models composed of zero-centered Lorentzians for band-limited noise (BLN) components, and Lorentzians for QPOs. 

\subsection{\xmm}
There are three \xmm\ ToO observations of \max15. Details of the observations are given in Table \ref{Tab:Obs_xmm}. We filtered and extracted the pn event file, using standard SAS (version 14.0.0) tools, paying particular attention to extract the list of photons not randomized in time. As the timing mode data, which are only available for the first observation, are affected by numerous short gaps, we only use the burst mode data in our analysis. Using the SAS task \texttt{epatplot} we made sure that the burst mode data are not affected by pile-up, by checking that the observed pattern distribution of the selected events follows the theoretical prediction. We then selected source photons from a stripe of 13 columns centred on the column with the highest count rate. We included single and double events (PATTERN$\le$4) in our PDS and energy spectra. For the PDS, covering the 1 -- 10 keV range, the contribution due to Poissonian noise was subtracted and the normalised PDS were converted to square fractional rms. We extracted energy spectra and corresponding background spectra ($3 \le$ RAWX $\le 5$), redistribution matrices, and ancillary response files for all observations.  

\begin{table*}
\caption{Details of \xmm\ observations}
\begin{center}
\begin{tabular}{lrlrrrr}
\hline\noalign{\smallskip}
 \multicolumn{1}{c}{\#} & \multicolumn{1}{c}{Obs.~id.} & \multicolumn{1}{c}{Date}  & \multicolumn{1}{c}{Mode$^{a}$} &  \multicolumn{1}{c}{Net Exp. [ks]}  &  \multicolumn{1}{c}{Exp.$^{b}$ [ks]} \\
 \hline\noalign{\smallskip}
1 & 0795711801  & 2017-09-07 & T/B & 34.0 & 19.1$^{c}$ \\
\noalign{\smallskip}
2 & 0795712001 & 2017-09-14 & B & 27.4 & 27.4 \\
\noalign{\smallskip}
3 & 0795712101 & 2017-09-15& B & 15.6 & 12.1 \\
\noalign{\smallskip}
\hline\noalign{\smallskip} 
\end{tabular} 
\end{center}
Notes: \\
$^{a}$: T for timing mode, B for burst mode\\ 
$^{b}$: longest interval of continuous exposure\\
$^{c}$: longest exposure in burst mode 
\label{Tab:Obs_xmm}
\end{table*}

\subsection{NICER}
The Neutron star Interior Composition Explorer \citep[NICER;][]{2012SPIE.8443E..13G} observed \max15\ between 2017 September 7th and October 11th, where the first three observations have exposures of less than 1 ks, and have hence been excluded from our study. These are the NICER observations taken during the time period in which the \swift/XRT monitoring observations reported here (Sect.~\ref{Sec:obs_sw}) have been obtained. NICER resumed observing \max15\ in January 2018, but these observations will not be included in this study. We made use of the pre-processed event files provided by the NICER datacenter and used these files to derive Poissonian noise subtracted, Leahy normalised and to square fractional rms converted  PDS in the 0.2 -- 10 keV range. We do not study energy spectra, as the averaged spectral response file, the only one available at the time, does not allow us to obtain robust spectral parameters and the fitted spectra are still affected by strong residuals. See however, \citet{2018ApJ...860L..28M} for a spectral study of one NICER observation of \max15.

\begin{figure}
\resizebox{\hsize}{!}{\includegraphics[clip,angle=0]{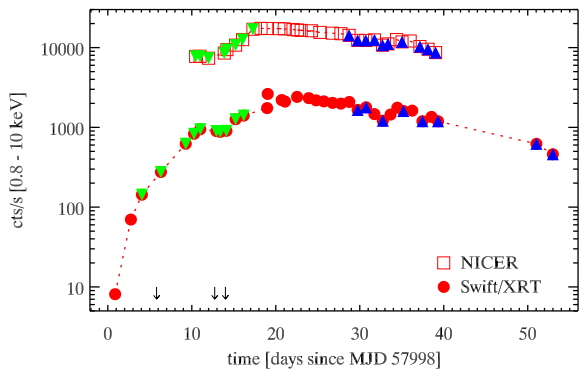}}
\caption{Light curve of \max15\ in fall 2017. Each data point represents one observation. Observations in which a type-C QPO has been detected are marked by triangles. Down-pointing (green) triangles indicate observations with type-C QPO when the source was brightening, while upward (blue) triangles indicate observations taken after the brightest observation. Arrows indicate the time of \xmm\ observations. T=0 corresponds to September 2nd 2017 00:00:00.000 UTC.}
\label{Fig:LC}
\end{figure}

\begin{figure}
\resizebox{\hsize}{!}{\includegraphics[clip,angle=0]{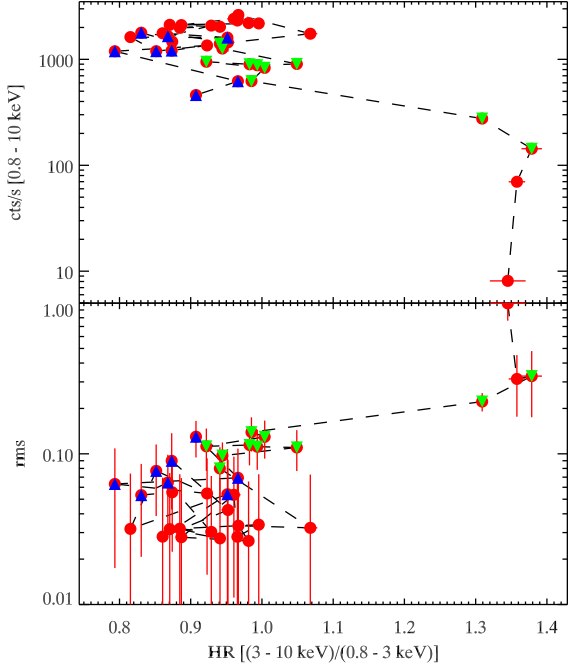}}
\caption{Hardness-intensity diagram (upper panel) and hardness-rms diagram (lower panel), derived using \swift/XRT count rates. Each data point represents one observation. Symbols are the same as in Fig.~\ref{Fig:LC}.}
\label{Fig:HID}
\end{figure}

\begin{figure}
\resizebox{\hsize}{!}{\includegraphics[clip,angle=0]{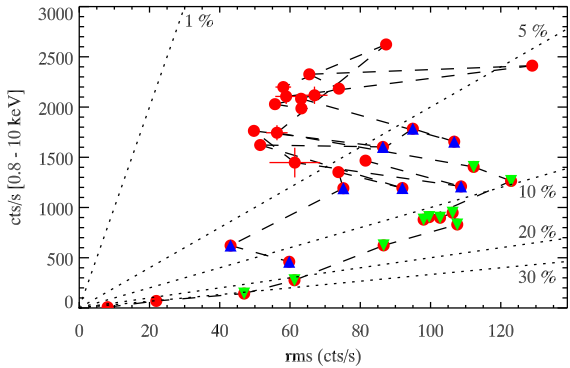}}
\caption{rms-intensity diagram, derived using \swift/XRT count rates. Each data point represents one observation. Symbols are the same as in Fig.~\ref{Fig:LC}.}
\label{Fig:RID}
\end{figure}
\section[]{Results}
\label{Sec:res}

\subsection{Diagnostic diagrams}
\label{Sec:diag}
Based on \swift/XRT data we determined source count rates in the total (0.8 -- 10 keV), soft (0.8 -- 3 keV), and hard (3 -- 10 keV) energy bands. We also derived NICER count rates in the 0.8 -- 10 keV band. The \swift/XRT and NICER light curves are shown in Fig.~\ref{Fig:LC}. Hardness ratios (HR) are derived by dividing the \swift/XRT count rate observed in the hard band by the one obtained in the soft band. We determined the fractional rms in the 0.3 -- 10 keV band and in the $4\times10^{-3}$ -- 35.13 Hz frequency range (with the exception of observation 00771371000 where we used the 0.02 -- 35.13 Hz range). The HID and HRD of the 2017 outburst of \max15\ are shown in Fig.~\ref{Fig:HID}, while Fig.~\ref{Fig:RID} is the RID. 

After the detection of the outburst the source increases in brightness. Exceeding a \swift/XRT count rate of $\sim$200 cts/s the source starts to soften. In the observation on day 9.3 the HR jumps to 1.0.\@ The source then further increases in intensity and remains softer than 1.1.\@ With increasing intensity the fractional rms decreases as can be seen in the HRD. After the highest intensity has been reached the source continues to soften, and the intensity decreases at softer HRs. During this decrease type-C QPOs are detectable (see Sect.~\ref{Sec:time}). The shape of this HID differs from the q-shape observed in many black hole X-ray binaries, where the soft state observations are observed at the softest HRs and observations with type-C QPOs are observed at harder HRs \citep[\eg\ ][]{2011MNRAS.418.2292M,2016MNRAS.460.1946S}. 

The RID shows the hard line at outburst rise, which lies at lower fractional rms values as seen \eg\ in \gx339\ \citep{2011MNRAS.410..679M}. In this diagram the soft state is more obvious than in the HID and  observations with QPOs observed after reaching the highest intensity correspond with excursions towards higher rms values. Based on the rms values, \max15\ seems to enter the soft state (rms below 5\%) on day 18 and to leave it on day 36 with an excursion to higher rms ratios around day 31.\@   

\subsection{Spectral properties}
\label{Sec:spec}
\subsubsection{\swift}

We used \textsc{Xspec} \citep[V.\ 12.8.2;][]{1996ASPC..101...17A} to fit the energy spectra in the 0.6 -- 10 keV range. Softer energies (below 0.6 keV) are omitted as the spectra are affected by a turn-up in this energy range, which is due to RMF redistribution modelling issues\footnote{http://www.swift.ac.uk/analysis/xrt/digest\_cal.php}. 
We grouped spectra to contain at least 20 counts in each bin to use $\chi^2$ minimisation for obtaining the best fit. We fitted the observed spectra with different one-component models, including foreground absorption \citep[\texttt{tbabs};][]{2000ApJ...542..914W}, using the abundances of \citet{2000ApJ...542..914W} and the cross sections given in \citet{1996ApJ...465..487V}. A power law model gives statistically unacceptable fits, with high photon indices and foreground absorption, when \max15\ is in the soft state (observations taken between day 18 and day 35). For these observations statistically acceptable fits can be obtained using an absorbed \texttt{diskbb} model \citep{1984PASJ...36..741M}, which supports their soft state nature.

\begin{figure}
\centering
\resizebox{\hsize}{!}{\includegraphics[clip,angle=0]{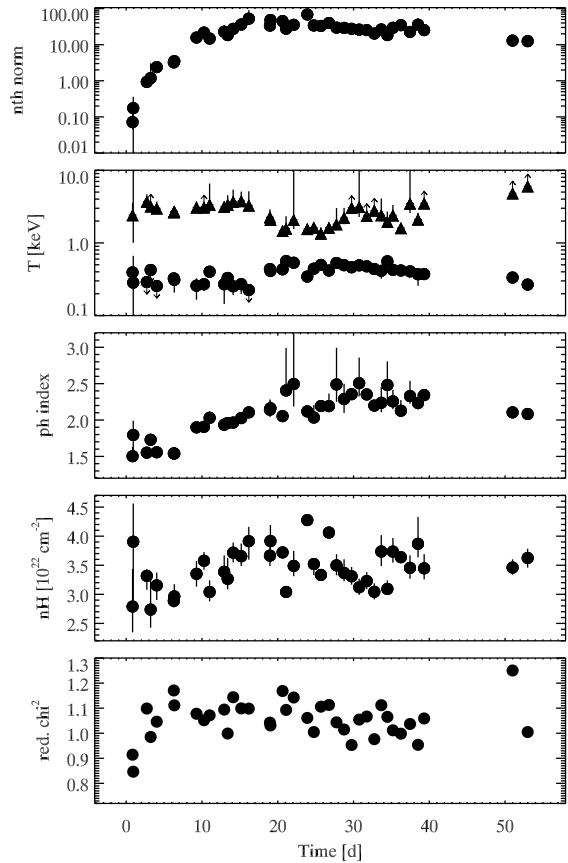}}
\caption{Evolution of spectral parameters, fitting \swift/XRT spectra with an absorbed thermal Comptonization model. Given parameters are (from bottom to top): reduced $\chi^2$, foreground absorption, photon index, inner disk and electron temperature (filled triangles), and \texttt{nthcomp} normalization. Arrows indicate upper limits on the disk temperature and lower limits on the electron temperature.}
\label{Fig:spec}
\end{figure}

Using an absorbed thermal Comptonization model \citep[\texttt{nthcomp}][]{1996MNRAS.283..193Z,1999MNRAS.309..561Z} we obtain statistically good fits with reduced $\chi^2$ values around 1 (0.85 -- 1.25) for all observations. Individual spectral parameters and reduced $\chi^2$ values are given in Table \ref{Tab:spec_par_nth} and the evolution of the spectral parameters is shown in Fig.~\ref{Fig:spec}. The foreground absorption ranges between 2.7 and 4.3\hcm{22}. The photon index increases with increasing intensity and ranges between 1.5 and 2.5.\@  The disk temperature is rather low between 0.2 and 0.6 keV, and for the first few observations only upper limits on the disk temperature can be obtained. The averaged disk temperature in the observations up to day 18, when the source shows type-C QPOs (see Sect.~\ref{Sec:time}), is $0.30\pm0.03$ keV, while the averaged disk temperature of observations taken between day 18 and 35 is $0.476_{-0.05}^{+0.04}$ keV. This increase in temperature is in agreement with a transition to the soft state. The electron temperatures obtained are low with values between 1.3 and 6.0 keV, and for some observations only a lower limit can be found. It is a known issue that spectra which only cover energies up to 10 keV give low electron temperatures. We added a \texttt{cflux} component to our best-fitting model to derive fluxes with errors. The obtained absorbed fluxes in the 0.6 -- 10 keV band range between $1.05\pm0.02$\ergcm{-9} and $3.71\pm0.02$\ergcm{-7}, making it one of the brightest black hole X-ray binary candidates ever observed.

Given the short exposures of \swift/XRT observations and the limited energy range and spectral resolution of these observations (compared to \eg\ \nus\ or NICER) using more complex, multi-component models will not give us additional insights, as the data will not allow us to disentangle contributions form different spectral components and degeneracies between different components will lead to increased uncertainties of spectral parameters.  

\subsubsection{\xmm}
We fitted the \xmm\ spectra in the 1 -- 10 keV range ignoring energies between 1.8 and 2.4 keV as this energy range shows features caused by gain shift owing to charge transfer inefficiency \citep{2011MNRAS.411..137H,2014A&A...571A..76D}. Energies below 1 keV are excluded as it is known that energy spectra obtained from EPIC/pn fast-readout mode data show excess emission at these soft energies \citep[see \eg\ ][]{2006A&A...448..677M}. 
Using an absorbed thermal Comptonization model, like we did for the \swift/XRT data, gives statistically unacceptable fits that show residuals at the soft end of the energy range and around 6 -- 7 keV. Thus we included a disk blackbody component and a Gaussian to improve the fits. Individual spectral parameters and reduced $\chi^2$ values can be found in Table \ref{Tab:spec_par_xmm}. The obtained foreground absorption and photon indices agree with values from the \swift/XRT spectra taken around the \xmm\ ones. The obtained inner disk temperatures are very low and the inner disk radius are very large, consistent with an accretion disk being truncated far away from the black hole, which is what one would expect for the LHS in the truncated disk geometry \citep{2007A&ARv..15....1D,2010LNP...794...53B,2014ARA&A..52..529Y}.   

\begin{table*}
\caption{Parameters with their 90\% error range of the absorbed thermal Comptonization model fitted to \swift/XRT spectra (0.6 -- 10 keV)}
\begin{center}
\begin{tabular}{rrlllllr}
\hline\noalign{\smallskip}
 \multicolumn{1}{c}{day} & \multicolumn{1}{c}{$\chi^2_{\mr{red}}\ (dof)$} & \multicolumn{1}{c}{\nh} & \multicolumn{1}{c}{$\Gamma$} & \multicolumn{1}{c}{$kT_{\mr{e}}$} & \multicolumn{1}{c}{$T_0$}& \multicolumn{1}{c}{norm}& \multicolumn{1}{c}{$F_{\mr{0.6-10keV}}^{\mr{abs}}$}  \\
 \multicolumn{1}{c}{} & \multicolumn{1}{c}{} & \multicolumn{1}{c}{\hcm{22}} & \multicolumn{1}{c}{} & \multicolumn{1}{c}{keV} & \multicolumn{1}{c}{eV}& \multicolumn{1}{c}{}& \multicolumn{1}{c}{erg cm$^{-2}$ s$^{-1}$}  \\
\hline\noalign{\smallskip}
0.84  &0.91 (421)  &$2.79  _{-0.45  }^{+0.65}  $&$1.50  _{-0.06  }^{+0.13}  $&$2.41  _{-0.33  }^{+1.15}  $&$<394    $&$0.07    _{-0.03  }^{+0.08}  $&$  1.05E-09\pm  1.61E-11$\\
\smallskip
 2.72  &1.10 (689) &$3.32  _{-0.24  }^{+0.12}  $&$1.55  _{-0.02  }^{+0.02}  $&$3.68  _{-0.49  }^{+0.98}  $&$<290    $&$0.93    _{-0.22  }^{+0.41}  $&$  9.95E-09\pm  6.95E-11$\\
\smallskip
 3.22  &0.99 (360) &$2.74  _{-0.31  }^{+0.62}  $&$1.73  _{-0.05  }^{+0.07}  $&$>3.18 $&$<425    $&$1.19    _{-0.31  }^{+1.86}  $&$  1.38E-08\pm  2.34E-10$\\
\smallskip
 4.03  &1.05 (514) &$3.15  _{-0.25  }^{+0.22}  $&$1.56  _{-0.04  }^{+0.04}  $&$2.98  _{-0.40  }^{+0.72}  $&$<255    $&$2.40    _{-0.61  }^{+0.84}  $&$  2.35E-08\pm  2.94E-10$\\
\smallskip
 6.28  &1.17 (828) &$2.89  _{-0.11  }^{+0.15}  $&$1.54  _{-0.01  }^{+0.01}  $&$2.64  _{-0.10  }^{+0.11}  $&$325  _{-65  }^{+46}  $&$3.24    _{-0.40  }^{+0.70}  $&$  3.79E-08\pm  1.29E-10$\\
\smallskip
 6.34  &1.11 (773) &$2.96  _{-0.16  }^{+0.21}  $&$1.54  _{-0.02  }^{+0.02}  $&$2.71  _{-0.14  }^{+0.17}  $&$310  _{-103  }^{+62}  $&$3.50    _{-0.58  }^{+1.18}  $&$  3.98E-08\pm  1.95E-10$\\
\smallskip
 9.27  &1.08 (702) &$3.35  _{-0.22  }^{+0.22}  $&$1.90  _{-0.03  }^{+0.04}  $&$3.08  _{-0.36  }^{+0.64}  $&$258  _{-93  }^{+67}  $&$15.78  _{-3.66  }^{+6.27}  $&$  9.55E-08\pm  5.53E-10$\\
\smallskip
 10.26  &1.05 (680) &$3.58  _{-0.15  }^{+0.15}  $&$1.91  _{-0.03  }^{+0.02}  $&$>3.08  $&$270  _{-52  }^{+44}  $&$21.66  _{-3.50  }^{+4.70}  $&$  1.43E-07\pm  9.33E-10$\\
\smallskip
 11.01  &1.07 (669) &$3.04  _{-0.16  }^{+0.20}  $&$2.03  _{-0.06  }^{+0.09}  $&$3.36  _{-0.70  }^{+3.19}  $&$402  _{-58  }^{+50}  $&$14.79  _{-2.15  }^{+3.38}  $&$  1.37E-07\pm  7.55E-10$\\
\smallskip
 12.93  &1.09 (711) &$3.39  _{-0.21  }^{+0.28}  $&$1.94  _{-0.03  }^{+0.05}  $&$3.14  _{-0.39  }^{+0.42}  $&$270  _{-125  }^{+62}  $&$22.86  _{-4.99  }^{+12.34}  $&$  1.19E-07\pm  7.91E-10$\\
\smallskip
 13.39  &1.00 (701) &$3.27  _{-0.18  }^{+0.36}  $&$1.96  _{-0.03  }^{+0.04}  $&$3.35  _{-0.51  }^{+1.18}  $&$330  _{-83  }^{+52}  $&$18.44  _{-3.15  }^{+7.34}  $&$  1.28E-07\pm  7.17E-10$\\
\smallskip
 14.12  &1.14 (708) &$3.72  _{-0.17  }^{+0.17}  $&$1.96  _{-0.04  }^{+0.04}  $&$3.68  _{-0.60  }^{+1.71}  $&$253  _{-61  }^{+47}  $&$27.43  _{-4.94  }^{+7.30}  $&$  1.55E-07\pm  8.70E-10$\\
\smallskip
 15.18  &1.10 (757) &$3.65  _{-0.17  }^{+0.22}  $&$2.03  _{-0.03  }^{+0.04}  $&$3.76  _{-0.56  }^{+1.32}  $&$271  _{-72  }^{+46}  $&$36.21  _{-6.46  }^{+12.79}  $&$  2.39E-07\pm  1.35E-09$\\
\smallskip
 16.18  &1.10 (692) &$3.91  _{-0.32  }^{+0.25}  $&$2.10  _{-0.06  }^{+0.08}  $&$3.27  _{-0.54  }^{+1.86}  $&$<225    $&$52.53  _{-17.10}^{+37.79}  $&$  2.02E-07\pm  8.95E-10$\\
\smallskip
 18.97  &1.04 (670) &$3.66  _{-0.18  }^{+0.22}  $&$2.14  _{-0.10  }^{+0.14}  $&$2.25  _{-0.30  }^{+0.64}  $&$436  _{-63  }^{+54}  $&$33.80  _{-4.92  }^{+8.04}  $&$  2.62E-07\pm  1.54E-09$\\
\smallskip
 19.03  &1.03 (681) &$3.92  _{-0.18  }^{+0.28}  $&$2.16  _{-0.10  }^{+0.10}  $&$2.08  _{-0.25  }^{+0.33}  $&$412  _{-74  }^{+48}  $&$48.17  _{-7.13  }^{+15.23}  $&$  3.36E-07\pm  1.94E-09$\\
\smallskip
 20.64  &1.17 (667) &$3.72  _{-0.03  }^{+0.03}  $&$2.05  _{-0.01  }^{+0.01}  $&$1.46  _{-0.10  }^{+0.01}  $&$433  \pm6  $&$44.98  _{-0.23  }^{+0.23}  $&$  3.21E-07\pm  1.64E-09$\\
\smallskip
 21.10  &1.09 (616) &$3.04  _{-0.04  }^{+0.04}  $&$2.41  _{-0.02  }^{+0.58}  $&$1.57  _{-0.23  }^{+0.76}  $&$562  \pm6  $&$27.79  _{-0.17  }^{+0.17}  $&$  2.97E-07\pm  2.29E-09$\\
\smallskip
 22.10  &1.14 (586) &$3.49  _{-0.18  }^{+0.26}  $&$2.50  _{-0.31  }^{+0.69}  $&$2.07  _{-0.52  }^{+7.84}  $&$533  _{-86  }^{+69}  $&$35.52  _{-4.75  }^{+9.47}  $&$  3.71E-07\pm  2.22E-09$\\
\smallskip
 23.90  &1.06 (639) &$4.28  _{-0.04  }^{+0.04}  $&$2.12  _{-0.01  }^{+0.01}  $&$1.55  _{-0.10  }^{+0.16}  $&$345  \pm8  $&$68.67  _{-17.72}^{+0.41}  $&$  2.64E-07\pm  1.62E-09$\\
\smallskip
 24.76  &1.00 (652) &$3.52  _{-0.19  }^{+0.10}  $&$2.03  _{-0.02  }^{+0.17}  $&$1.62  _{-0.15  }^{+0.23}  $&$442  _{-34  }^{+67}  $&$34.05  _{-5.35  }^{+3.03}  $&$  2.51E-07\pm  1.57E-09$\\
\smallskip
 25.69  &1.11 (612) &$3.33  _{-0.04  }^{+0.04}  $&$2.19  _{-0.02  }^{+0.02}  $&$1.34  _{-0.13  }^{+0.29}  $&$498  \pm6  $&$33.04  _{-0.21  }^{+0.21}  $&$  2.58E-07\pm  1.49E-09$\\
\smallskip
 26.75  &1.11 (665) &$4.06  _{-0.04  }^{+0.04}  $&$2.19  _{-0.01  }^{+0.17}  $&$1.61  _{-0.15  }^{+0.24}  $&$417  _{-7  }^{+60}  $&$39.92  _{-6.67  }^{+0.23}  $&$  2.38E-07\pm  1.64E-09$\\
\smallskip
 27.75  &1.04 (621) &$3.50  _{-0.17  }^{+0.19}  $&$2.49  _{-0.24  }^{+0.50}  $&$1.81  _{-0.31  }^{+1.21}  $&$530  _{-44  }^{+58}  $&$29.69  _{-3.59  }^{+4.31}  $&$  2.43E-07\pm  1.56E-09$\\
\smallskip
 28.75  &1.02 (662) &$3.37  _{-0.14  }^{+0.23}  $&$2.29  _{-0.19  }^{+0.21}  $&$2.21  _{-0.41  }^{+0.83}  $&$494  _{-74  }^{+47}  $&$29.23  _{-3.30  }^{+7.03}  $&$  2.11E-07\pm  1.43E-09$\\
\smallskip
 29.74  &0.95 (656) &$3.31  _{-0.14  }^{+0.16}  $&$2.35  _{-0.04  }^{+0.05}  $&$>3.03 $&$462  _{-34  }^{+32}  $&$27.53  _{-2.99  }^{+3.90}  $&$  2.43E-07\pm  1.52E-09$\\
\smallskip
 30.74  &1.05 (662) &$3.12  _{-0.13  }^{+0.15}  $&$2.51  _{-0.18  }^{+0.35}  $&$3.12  _{-0.85  }^{+38.34}  $&$491  \pm43  $&$26.17  _{-2.67  }^{+3.48}  $&$  2.04E-07\pm  1.25E-09$\\
\smallskip
 31.74  &1.07 (629) &$3.23  _{-0.14  }^{+0.15}  $&$2.35  _{-0.05  }^{+0.05}  $&$>2.35 $&$477  _{-34  }^{+32}  $&$25.43  _{-2.68  }^{+3.42}  $&$  1.97E-07\pm  1.40E-09$\\
\smallskip
 32.74  &0.98 (677) &$3.04  _{-0.13  }^{+0.15}  $&$2.20  _{-0.03  }^{+0.04}  $&$>2.79  $&$438  _{-34  }^{+31}  $&$20.38  _{-2.15  }^{+2.82}  $&$  1.57E-07\pm  9.82E-10$\\
\smallskip
 33.66  &1.11 (605) &$3.74  _{-0.25  }^{+0.28}  $&$2.24  _{-0.13  }^{+0.22}  $&$2.40  _{-0.44  }^{+1.69}  $&$413  _{-92  }^{+69}  $&$26.42  _{-5.22  }^{+11.39}  $&$  1.81E-07\pm  1.47E-09$\\
\smallskip
 34.46  &1.07 (669) &$3.09  _{-0.10  }^{+0.15}  $&$2.48  _{-0.27  }^{+0.32}  $&$1.95  _{-0.35  }^{+0.75}  $&$559  _{-59  }^{+41}  $&$18.48  _{-1.43  }^{+2.55}  $&$  1.65E-07\pm  9.00E-10$\\
\smallskip
 35.19  &1.01(676)  &$3.73  _{-0.19  }^{+0.23}  $&$2.26  _{-0.11  }^{+0.16}  $&$2.39  _{-0.37  }^{+0.93}  $&$426  _{-59  }^{+52}  $&$29.41  _{-4.47  }^{+7.18}  $&$  2.07E-07\pm  1.24E-09$\\
\smallskip
 36.25  &1.00 (657) &$3.64  _{-0.04  }^{+0.04}  $&$2.13  _{-0.01  }^{+0.15}  $&$1.60  _{-0.13  }^{+0.20}  $&$418  \pm7  $&$34.68  _{-3.35  }^{+0.20}  $&$  2.33E-07\pm  1.34E-09$\\
\smallskip
 37.45  &1.04 (673) &$3.45  _{-0.19  }^{+0.22}  $&$2.33  _{-0.10  }^{+0.21}  $&$3.48  _{-0.90  }^{+23.82}  $&$406  _{-51  }^{+49}  $&$22.61  _{-3.41  }^{+4.96}  $&$  1.50E-07\pm  9.24E-10$\\
\smallskip
 38.51  &0.95 (647) &$3.87  _{-0.23  }^{+0.46}  $&$2.23  _{-0.09  }^{+0.13}  $&$2.10  _{-0.26  }^{+0.50}  $&$371  _{-114  }^{+60}  $&$35.79  _{-7.07  }^{+17.59}  $&$  2.10E-07\pm  1.33E-09$\\
\smallskip
 39.32  &1.06 (625) &$3.45  _{-0.19  }^{+0.24}  $&$2.34  _{-0.03  }^{+0.04}  $&$>3.50  $&$372  _{-51  }^{+39}  $&$25.35  _{-3.98  }^{+6.87}  $&$  1.50E-07\pm  1.10E-09$\\
\smallskip
 50.99  &1.25 (768) &$3.46  _{-0.12  }^{+0.14}  $&$2.11  _{-0.01  }^{+0.01}  $&$>4.81  $&$334  _{-35  }^{+29}  $&$12.98  _{-1.51  }^{+2.16}  $&$  8.21E-08\pm  3.24E-10$\\
\smallskip
 52.98  &1.00 (729) &$3.62  _{-0.16  }^{+0.16}  $&$2.08  _{-0.02  }^{+0.02}  $&$>6.00  $&$267  _{-50  }^{+43}  $&$12.51  _{-2.17  }^{+2.96}  $&$  6.68E-08\pm  3.42E-10$\\
\hline\noalign{\smallskip}
\end{tabular} 
\end{center}
\label{Tab:spec_par_nth}
\end{table*}

\begin{table*}
\caption{Parameters with their 90\% error range of the absorbed thermal Comptonization model fitted to \xmm\ spectra (1 -- 1.8 \& 2.4 -- 10 keV)}
\begin{center}
\begin{tabular}{rrlllll}
\hline\noalign{\smallskip}
 \multicolumn{1}{c}{day} & \multicolumn{1}{c}{$\chi^2_{\mr{red}}\ (dof)$} & \multicolumn{1}{c}{\nh} & \multicolumn{1}{c}{$\Gamma$} & \multicolumn{1}{c}{$kT_{\mr{e}}$} & \multicolumn{1}{c}{$T_0$}& \multicolumn{1}{c}{norm}  \\
 \multicolumn{1}{c}{} & \multicolumn{1}{c}{} & \multicolumn{1}{c}{\hcm{22}} & \multicolumn{1}{c}{} & \multicolumn{1}{c}{keV} & \multicolumn{1}{c}{eV}& \multicolumn{1}{c}{}  \\
\hline\noalign{\smallskip}
5.81 & 1.07 (1674) & $3.219_{-0.017}^{+0.005}$ & $1.526_{-0.011}^{+0.012}$ & $>899.77$ & $74.4_{-0.4}^{+0.6}$ & $3.13_{-0.04}^{+0.05}$  \\
\smallskip
12.73 & 1.15 (1674) & $3.692_{-0.012}^{+0.013}$ & $2.025_{-0.026}^{+0.008}$ & $7.22_{-0.59}^{+1.21}$ & $86.1_{-0.3}^{+0.2}$ & $19.20_{-0.16}^{+0.15}$ \\
\smallskip
14.00 & 1.08 (1674) & $3.595_{-0.017}^{+0.015}$ & $1.970_{-0.002}^{+0.005}$ & $6.08_{-0.52}^{+0.73}$ & $85.2\pm0.4$ & $20.79_{-0.05}^{+0.01}$ \\
\hline\noalign{\smallskip}
\end{tabular} 
\end{center}
\label{Tab:spec_par_xmm}
\end{table*}

\subsection{Timing properties}
\label{Sec:time}
\subsubsection{\swift}
\label{Sec:time_sw}
\begin{figure*}
\resizebox{\hsize}{!}{\includegraphics[clip,angle=0]{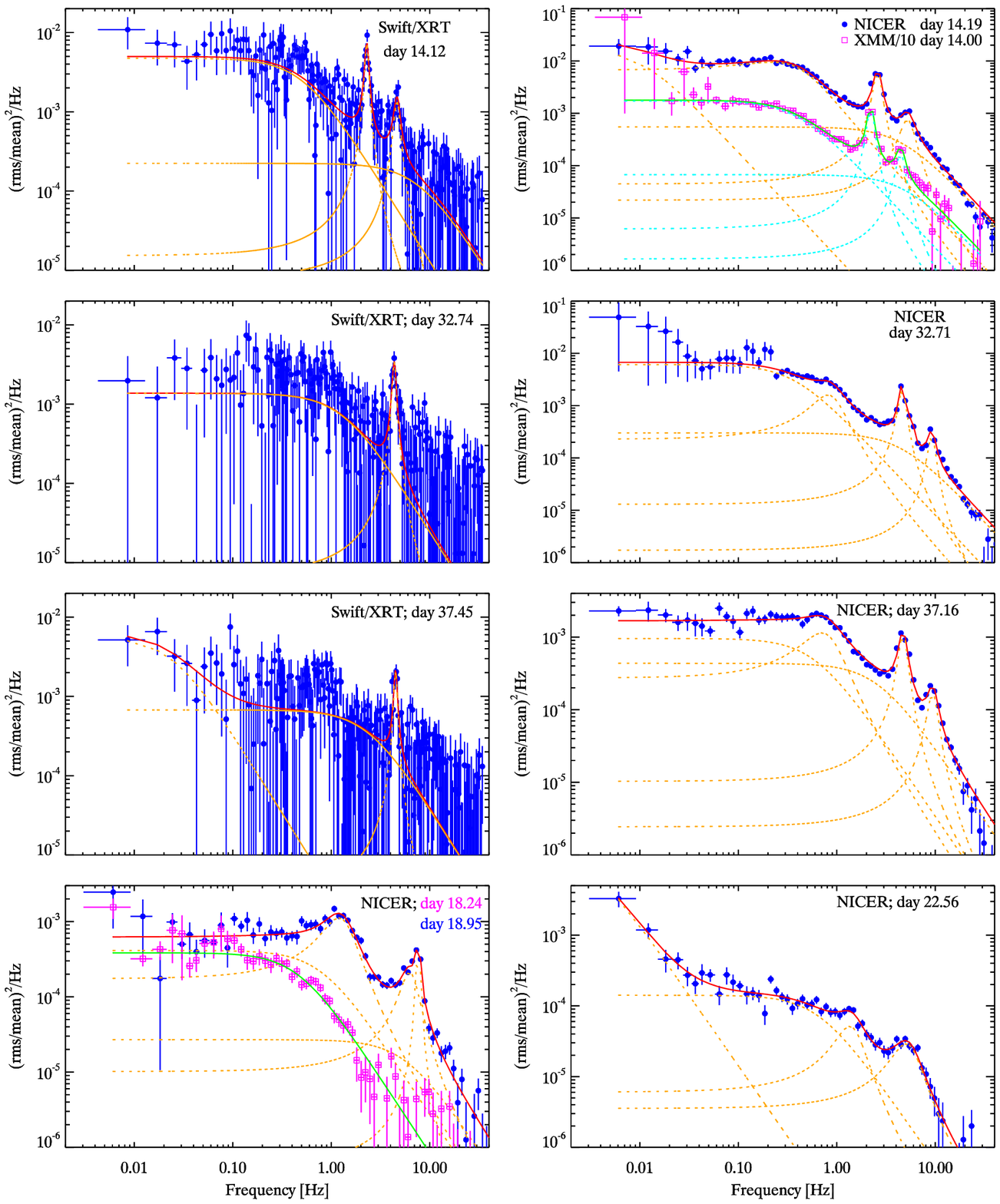}}
\caption{Examples of PDS for three \swift/XRT observations, one taken during brightening and two during fainting of the outburst. PDS of three NICER and one \xmm\ observations taken on the same day as the \swift/XRT observations are shown. The \xmm\ data are scaled with a factor of 0.1 to enhance clarity of the plot. In addition, two NICER PDS obtained on day 18 and when the source was in the soft state are shown.}
\label{Fig:pds}
\end{figure*}

The PDS of the first 10 observations with a count rate above 100 cts/s show two BLN components and a low frequency QPO (Fig.~\ref{Fig:pds}). The characteristic frequency of the QPO \citep[$\nu_{\mr{max}}=\sqrt{\nu_0^2+\Delta^2}$, where $\nu_0$ is the centroid frequency, and $\Delta$ is the half width at half maximum][]{2002ApJ...572..392B} increases from 0.17 to 3.22 Hz. Apart from the first two and the last one of these observations, an upper harmonic of the QPO is present. Some observations also show a lower harmonic or additional peaked noise. Details on the BLN and the QPOs can be found in Tables~\ref{Tab:pds1} and \ref{Tab:pds2}, respectively.

In the then following observations the PDS are dominated by power law noise. For observations taken after the outburst reached its maximum, the PDS of those observations with an rms value above 6\% show again at least one BLN component and a QPO (Fig.~\ref{Fig:pds}). A QPO with $\sigma>3$ is also detected in the observation taken on day 30.74, which has a total fractional rms value of $\sim5.3$\%, and in the observation taken on day 35.19, which has a total fractional rms value of $\sim5.4$\%, a QPO with $\sigma=2.8$ is detected. In these observations the characteristic frequency of the QPO is higher (2.51 -- 6.48 Hz) and no harmonics are present. The first seven of these observations (between day 29.74 and 50.99) have a total fractional rms in the range of 5 -- 10\%, the rms range in which type-B QPOs are observed, although the error bars are quite big and rms values above 10\% are within the error range. These observations comprise the observations that require only one BLN component and the characteristic frequency of this BLN component is always lower than the characteristic frequency of the QPO (see Tables~\ref{Tab:pds1} and \ref{Tab:pds2} and Fig.~\ref{Fig:pds}), which means that the QPO is sitting on the decaying part of the noise component. In all other observations with a QPO we observe at least one noise component with a characteristic frequency above that of the QPO. This is also the case for the last observation with a QPO, which as a total rms value of $\sim12$\%. This oscillation is for sure a type-C QPO, and the detection of radio emission around the time of this observation \citep{2017ATel10899....1R} gives additional evidence that \max15\ was in a hard state at the time of this observation. 

\begin{figure}
\resizebox{\hsize}{!}{\includegraphics[clip,angle=0]{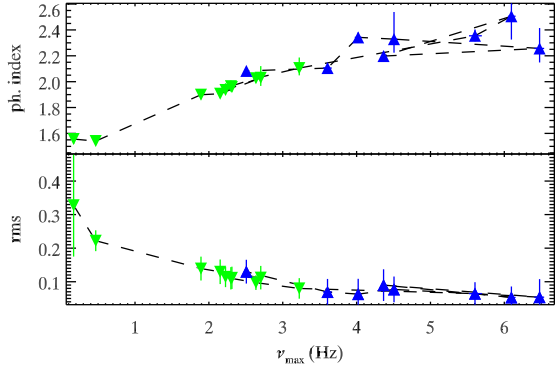}}
\caption{Correlation of the photon index obtained using the thermal Comptonization model (upper panel) and anti-correlation of the total fractional rms variability (lower panel) with the characteristic QPO frequency. Down-pointing (green) triangles indicate observations with a type-C QPO when the source is brightening, while upward (blue) triangles indicate observations taken after the brightest one.}
\label{Fig:fchar_rms}
\end{figure}

\begin{figure}
\resizebox{\hsize}{!}{\includegraphics[clip,angle=0]{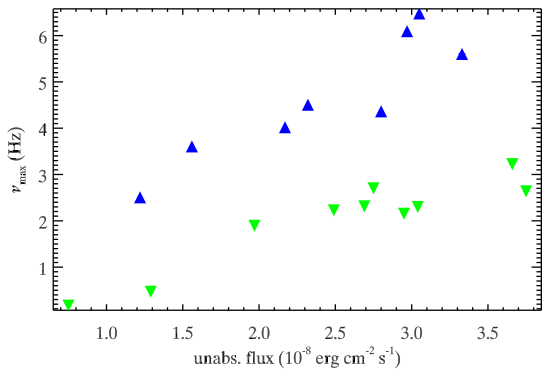}}
\caption{Correlations of the characteristic QPO frequency and the unabsorbed source flux. (Symbols as in Fig.~\ref{Fig:fchar_rms}.)}
\label{Fig:fchar_fl}
\end{figure}

We find an anti-correlation of the characteristic QPO frequency and the total fractional rms variability (Fig.~\ref{Fig:fchar_rms}) for all QPOs, where the total fractional rms is $\ga10$\%. For the remaining seven observations  the correlation is flat. The anti-correlation is observed for type-C QPOs \citep{2011MNRAS.418.2292M}, and provides further evidence that the QPOs of \max15\ with an anti-correlation are of type-C. We find a linear correlation between the photon index (obtained using the \texttt{nthcomp} model; see Sect.~\ref{Sec:spec}) and the characteristic QPO frequency, which is quite similar to the correlations obtained for type-C QPOs in XTE\,J1650--500 and \h1743 \citep{2013MNRAS.429.2655S}. 
Regarding the dependence of the characteristic QPO frequency on the unabsorbed source flux (derived using the \texttt{nthcomp} model; see Sect.~\ref{Sec:spec}), we obtain two correlations for the QPOs observed during softening and hardening of the outburst (Fig.~\ref{Fig:fchar_fl}).

\begin{figure}
\resizebox{\hsize}{!}{\includegraphics[clip,angle=0]{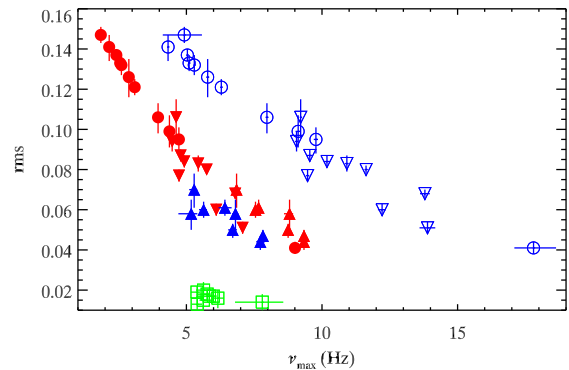}}
\caption{Correlation of the total fractional rms variability with the characteristic QPO frequency, derived from NICER data. Circles indicate observations taken during outburst rise, while down-pointing triangles indicate observations taken during outburst decay (open symbols indicate upper harmonics). Upward-pointing triangles indicate observations taken on days 18 and 19. Squares indicate observations with a flat correlation.}
\label{Fig:fchar_rms_ni}
\end{figure}

\begin{table*}
\caption{Parameters of the BLN components of the PDS}
\begin{center}
\begin{tabular}{rrrrr}
\hline\noalign{\smallskip}
 \multicolumn{1}{c}{day} & \multicolumn{1}{c}{$\nu_{\mr{max; BLN1}}$} & \multicolumn{1}{c}{rms$_{\mr{BLN1}}$} & \multicolumn{1}{c}{$\nu_{\mr{max; BLN2}}$} & \multicolumn{1}{c}{rms$_{\mr{BLN2}}$} \\
\hline\noalign{\smallskip}
 \multicolumn{5}{c}{\swift/XRT (0.3 -- 10 keV)}\\ 
\hline\noalign{\smallskip}
4.03   &$ 4.41_{-1.17}^{+0.59}  $&$ 0.170\pm0.017  $&$ 0.02_\pm0.01 $&$  0.088_{-0.023}^{+0.018} $\\
\smallskip
6.28   &$ 3.23_{-0.24}^{+0.33}  $&$ 0.139_{-0.006}^{+0.004}  $&$ 0.24_{-0.03}^{+0.04} $&$  0.159_{-0.009}^{+0.011} $\\
\smallskip
9.27  &$ 4.44_{-0.90}^{+0.56}  $&$ 0.096\pm0.007  $&$ 0.44_{-0.12}^{+0.11} $&$  0.061_{-0.010}^{+0.007} $\\
\smallskip
10.26   &$ 3.53_{-0.85}^{+1.47}  $&$ 0.081_{-0.011}^{+0.008}  $&$ 0.43_{-0.13}^{+0.16} $&$  0.052_{-0.009}^{+0.010} $\\
\smallskip
11.01   &$ 7.57_{-2.06}^{+2.43}  $&$ 0.064_{-0.014}^{+0.008}  $&$ 0.61_{-0.10}^{+0.12} $&$  0.062_{-0.005}^{+0.005} $\\
\smallskip
12.93   &$ 5.91_{-1.38}^{+1.64}  $&$ 0.071\pm0.006  $&$ 0.39_{-0.06}^{+0.05} $&$  0.058_{-0.004}^{+0.003} $\\
\smallskip
13.39   &$ 5.48_{-1.35}^{+1.54}  $&$ 0.079_{-0.006}^{+0.005}  $&$ 0.56_{-0.22}^{+0.17} $&$  0.052_{-0.009}^{+0.006} $\\
\smallskip
14.12   &$ 9.38_{-5.32}^{+0.62}  $&$ 0.052\pm0.010  $&$ 0.53_{-0.11}^{+0.08} $&$  0.063_{-0.007}^{+0.003} $\\
\smallskip
15.18   &$ 8.31_{-2.92}^{+1.69}  $&$ 0.048_{-0.008}^{+0.009}  $&$ 0.45_{-0.12}^{+0.05} $&$  0.058_{-0.010}^{+0.003} $\\
\smallskip
16.18  &$ 7.29_{-3.48}^{+2.71}  $&$ 0.032_{-0.011}^{+0.008}  $&$ 0.77_{-0.10}^{+0.11} $&$  0.051_{-0.004}^{+0.003} $\\
\smallskip
29.74 & $3.37_{- 1.02}^{+ 0.82}  $&$ 0.044\pm0.006  $&$ 0.005_{- 0.005}^{+ 0.006}  $&$ 	0.022_{- 0.006}^{+ 6.081}  $\\
\smallskip
30.74 & $2.12_{- 0.44}^{+ 0.45}  $&$ 0.039\pm0.003  $&$ 	--  $&$ 	-- $\\
\smallskip
32.74  &$ 1.35_{-0.21}^{+0.33}  $&$ 0.054\pm0.004  $&$ -- $&$  -- $\\
\smallskip
35.19  &$ 2.38_{-0.54}^{+0.74}  $&$ 0.043\pm0.005  $&$ -- $&$  -- $\\
\smallskip
37.45  &$ 2.42_{-0.71}^{+3.13}  $&$ 0.051_{-0.005}^{+6.053}  $&$ 0.02_{-0.01}^{+0.03} $&$  0.015\pm0.003 $\\
\smallskip
39.32  &$ 1.01_{-0.15}^{+0.18}  $&$ 0.052\pm0.003  $&$ -- $&$  -- $\\
\smallskip
50.99 &$ 0.64_{-0.14}^{+0.12}  $&$ 0.055_{-0.008}^{+0.006}  $&$ -- $&$  -- $\\
\smallskip
52.98 &$ 4.17_{-0.54}^{+0.70}  $&$ 0.106\pm0.005  $&$ 0.51_{-0.11}^{+0.10} $&$  0.061_{-0.007}^{+0.006} $\\
\hline\noalign{\smallskip}
 \multicolumn{5}{c}{\xmm\ (1 -- 10 keV)}\\ 
\hline\noalign{\smallskip}
5.81 & $4.974_{-0.648}^{+0.808}$ &  $0.158\pm 0.008$& $0.3092\pm0.012$ &  $0.2317\pm0.004$ \\
\smallskip
12.73 & $6.427_{-0.832}^{+1.053}$ & $0.068_{-0.005}^{+0.004}$&$0.484\pm0.014$&$0.103\pm0.001$ \\
\smallskip
14.00 & $4.934_{-0.946}^{+1.371} $& $0.072_{-0.010}^{+0.009}$&$0.395\pm0.024$&$0.103_{-0.003}^{+0.002} $\\
\hline\noalign{\smallskip}
\end{tabular} 
\end{center}
\label{Tab:pds1}
Notes:\\
rms: root mean square; $\nu_{\mr{max}}$: characteristic frequency; BLN: band limited noise
\end{table*}

\begin{table*}
\caption{Parameters of the noise components of the PDS derived from NICER data (0.2 -- 10 keV)}
\begin{center}
\footnotesize
\begin{tabular}{rrrrrrrrr}
\hline\noalign{\smallskip}
 \multicolumn{1}{c}{day} & \multicolumn{1}{c}{$\nu_{\mr{max; BLN1}}$} & \multicolumn{1}{c}{rms$_{\mr{BLN1}}$} & \multicolumn{1}{c}{$\nu_{\mr{max; BLN2}}$} & \multicolumn{1}{c}{rms$_{\mr{BLN2}}$} & \multicolumn{1}{c}{$\nu_{\mr{max; PN1}}$} & \multicolumn{1}{c}{rms$_{\mr{PN1}}$} & \multicolumn{1}{c}{Q$_{\mr{PN1}}$}  & \multicolumn{1}{c}{$\sigma_{\mr{PN1}}$}\\
\hline\noalign{\smallskip}
10.46& $1.261_{-0.492}^{+0.781} $ & $>0.041                  $ & $0.388_{-0.053}^{+0.038} $ & $0.079_{-0.012}^{+0.005} $ & $5.820_{-0.268}^{+0.389} $ & $0.060_{-0.005}^{+0.003} $ & 0.64 & 6.03 \\
\smallskip
10.99& $3.351_{-0.177}^{+0.279} $ & $0.070\pm 0.003          $ & $0.337_{-0.014}^{+0.017} $ & $0.083_{-0.001}^{+0.002} $ & -- & -- & -- &-- \\
\smallskip
12.00& $3.177_{-0.390}^{+0.853} $ & $0.059_{-0.015}^{+0.010} $ & $0.391_{-0.021}^{+0.023} $ & $0.091\pm 0.003          $ & $4.922_{-0.651}^{+0.793} $ & $0.055_{-0.012}^{+0.010} $ & 0.85 & 2.31\\
\smallskip
13.87& $3.342_{-0.162}^{+0.207} $ & $0.083_{-0.003}^{+0.002} $ &      --                      &              --              & $0.358_{-0.027}^{+0.029} $ & $0.079\pm 0.002          $ & 0.38 & 19.80\\
\smallskip
14.19& $4.445_{-0.302}^{+0.357} $ & $0.062\pm 0.003          $ & $0.010_{-0.006}^{+0.009} $ & $0.017_{-0.004}^{+0.003} $ & $0.407_{-0.012}^{+0.024} $ & $0.083_{-0.002}^{+0.001} $ & 0.32 & 27.77\\
\smallskip
15.22& $5.000_{-0.207}^{+0.000} $ & $0.043\pm 0.002          $ & $0.419_{-0.058}^{+0.034} $ & $0.067_{-0.008}^{+0.004} $ & $0.598_{-0.063}^{+0.056} $ & $0.044_{-0.006}^{+0.009} $ & 1.02 & 3.86\\
\smallskip
16.05& $4.292_{-0.390}^{+0.421} $ & $ 0.062\pm0.003 $ &-- &  --&$ 	0.471_{-0.037}^{+0.037} $ & $ 	0.075\pm	0.002	$ &	0.47&	16.37\\
\smallskip
16.38& $3.336_{-0.575}^{+0.459} $ & $ 0.044\pm0.003 $ &-- & -- &$ 	0.713_{-0.048}^{+0.045} $ & $ 	0.060\pm	0.002	$ &	0.56&	12.54\\
\smallskip
16.70& $4.139_{-0.516}^{+0.634} $ & $ 0.042\pm0.003 $ &-- & -- &$ 	0.611_{-0.042}^{+0.041} $ & $ 	0.069\pm	0.002	$ &	0.52&	17.23\\
\smallskip
16.83& $4.404_{-0.547}^{+0.666} $ & $ 0.040\pm0.003 $ &-- &-- &$  	0.765_{-0.040}^{+0.042} $ & $ 	0.057\pm	0.002	$ &	0.63&	16.76\\
\smallskip
17.28& $0.774_{-0.440}^{+0.630} $ & $>0.006                  $ & $0.005_{-0.003}^{+0.003} $ & $0.007\pm 0.001          $ & $6.981_{-0.285}^{+0.283} $ & $0.017_{-0.002}^{+0.001} $ & 2.06 & 5.16\\
\smallskip
18.24& $0.467_{-0.024}^{+0.025} $ & $0.017\pm 0.0003         $ & -- & -- &--  & -- &  --& --\\
\smallskip
18.95& $4.634_{-0.525}^{+0.618} $ & $0.031\pm 0.002          $ & $0.734_{-0.189}^{+0.308} $ & $0.021_{-0.003}^{+0.005} $ & $1.254\pm 0.031          $ & $0.035\pm 0.001 $ & 1.57 & 21.69\\
\smallskip
19.01& $4.709_{-0.616}^{+0.773} $ & $0.029_{-0.003}^{+0.002} $ & $0.941_{-0.286}^{+0.317} $ & $0.021\pm 0.005          $ & $1.327_{-0.038}^{+0.039} $ & $0.034\pm 0.001 $ & 1.38 & 24.36\\
\smallskip
19.08& $4.010_{-0.574}^{+0.801} $ & $0.028_{-0.004}^{+0.003} $ & $1.320_{-0.273}^{+0.321} $ & $0.023\pm 0.004          $ & $1.512_{-0.058}^{+0.056} $ & $0.023\pm 0.001 $ & 1.50 & 16.07\\
\smallskip
19.15& $3.292_{-0.554}^{+0.800} $ & $0.029_{-0.006}^{+0.004} $ & $1.213_{-0.491}^{+0.490} $ & $0.019_{-0.008}^{+0.007} $ & $1.510_{-0.045}^{+0.044} $ & $0.025\pm 0.001 $ & 1.57 & 18.14\\
\smallskip
19.21& $3.696_{-0.273}^{+0.294} $ & $0.029\pm 0.001          $ & $0.110_{-0.036}^{+0.046} $ & $0.008\pm 0.001          $ & $1.425_{-0.074}^{+0.075} $ & $0.025\pm 0.001 $ & 0.83 & 20.50\\
\smallskip
19.34& $6.097_{-1.027}^{+1.430} $ & $0.030_{-0.003}^{+0.002} $ &         --                 &             --                 & $1.257_{-0.064}^{+0.070} $ & $0.038\pm 0.001 $ & 1.07 & 14.46\\
\smallskip
19.41& $5.126_{-0.805}^{+0.902} $ & $0.037\pm 0.002          $ &            --              &           --                   & $1.135_{-0.055}^{+0.057} $ & $0.046\pm 0.001 $ & 1.10 & 16.25\\
\smallskip
19.85& $1.303_{-0.136}^{+0.162} $ & $0.014\pm 0.001          $ & $1.95e-5_{-8.81e-6}^{+0.002} $ & $0.084_{-0.010}^{+0.016} $ &--  & -- & --&-- \\
\smallskip
20.37& $1.368_{-0.116}^{+0.121} $ & $0.015\pm 0.0004         $ & $0.003_{-0.003}^{+0.002}$ & $0.006_{-0.001}^{+0.062} $ &--  &  --& -- &-- \\
\smallskip
21.01& $1.338_{-0.063}^{+0.074} $ & $0.016\pm 0.0002         $ & $0.015\pm 0.004            $ & $0.006_{-0.0003}^{+0.0003} $ & -- &  --& -- & --\\
\smallskip
22.05& $1.330_{-0.071}^{+0.074} $ & $>0.017                  $ & $0.009\pm 0.003            $ & $0.007\pm 0.001            $ & $0.203_{-0.005}^{+0.012} $ & $0.002\pm 0.001 $ & 7.89 & 2.00\\
\smallskip
22.57& $0.658_{-0.153}^{+0.336} $ & $0.012_{-0.001}^{+0.002} $ & $0.004\pm 0.002            $ & $>0.007                  $ &  --&  --& --& --\\
\smallskip
23.02& $1.483_{-0.074}^{+0.076} $ & $0.017\pm 0.0003         $ & $0.007\pm0.002$ & $0.006_{-0.0004}^{+0.0005} $ & -- &  --&--  & --\\
\smallskip
24.24& $0.654_{-0.036}^{+0.037} $ & $0.015\pm 0.0002         $ & $0.003_{-0.003}^{+0.003}$ & $0.006_{-0.001}^{+0.065} $ &  --& -- &--  & --\\
\smallskip
25.25& $0.993_{-0.111}^{+0.119} $ & $0.011\pm 0.0004         $ & $0.009\pm 0.002            $ & $0.006\pm 0.0003         $ & -- &  --& -- &-- \\
\smallskip
26.74& $1.656_{-0.128}^{+0.145} $ & $0.018\pm 0.001          $ & $0.007_{-0.003}^{+0.004} $ & $0.006\pm 0.001            $ & -- &--  &--  &-- \\
\smallskip
27.76& $1.393_{-0.128}^{+0.157} $ & $0.018\pm 0.0004         $ & $0.037_{-0.013}^{+0.018} $ & $0.007\pm 0.001          $ & -- & -- &--  &-- \\
\smallskip
28.73& $5.098_{-0.162}^{+0.172} $ & $0.019\pm 0.001          $ & $0.727_{-0.160}^{+0.185} $ & $0.025_{-0.003}^{+0.004} $ & $1.333_{-0.068}^{+0.061} $ & $0.030_{-0.003}^{+0.002} $ & 1.16 & 5.61\\
\smallskip
29.76& $4.212_{-0.243}^{+0.274} $ & $0.025\pm 0.004          $ & $1.052_{-0.188}^{+0.129} $ & $0.042_{-0.006}^{+0.004} $ & $0.947_{-0.047}^{+0.051} $ & $0.042_{-0.003}^{+0.004} $ & 1.08 & 7.07\\
\smallskip
30.72& $3.993\pm 0.089          $ & $0.023\pm 0.001          $ & $0.564_{-0.126}^{+0.109} $ & $0.028\pm 0.004          $ & $1.040_{-0.035}^{+0.032} $ & $0.050\pm 0.002 $ & 0.84 & 13.21\\
\smallskip
31.75& $5.011_{-0.142}^{+0.137} $ & $0.027_{-0.002}^{+0.001} $ & $0.568_{-0.122}^{+0.190} $ & $0.021_{-0.003}^{+0.004} $ & $1.215\pm 0.035          $ & $0.046_{-0.002}^{+0.001} $ & 0.91 & 12.18\\
\smallskip
32.71& $4.485_{-0.275}^{+0.405} $ & $0.046_{-0.003}^{+0.002} $ & $0.364_{-0.058}^{+0.082} $ & $0.059_{-0.003}^{+0.004} $ & $0.880_{-0.058}^{+0.059} $ & $0.038_{-0.005}^{+0.004} $ & 1.23 & 3.55\\
\smallskip
33.36& $2.239_{-0.592}^{+0.632} $ & $0.033_{-0.011}^{+0.006} $ & $0.711_{-0.206}^{+0.263} $ & $0.034_{-0.008}^{+0.011} $ & $0.945\pm 0.059          $ & $0.043_{-0.004}^{+0.005} $ & 0.85 & 5.33\\
\smallskip
34.38& $0.408_{-0.023}^{+0.024} $ & $0.015\pm 0.0003         $ &          --                  &          --                  & -- &--  & -- &-- \\
\smallskip
35.03& $0.716_{-0.051}^{+0.029} $ & $0.046\pm 0.001          $ &            --               &             --               & -- &--  &  --&-- \\
\smallskip
36.00& $0.453_{-0.024}^{+0.025} $ & $0.013\pm 0.0002 $ &-- &-- &-- &-- &-- &-- \\ 
\smallskip
37.16& $2.633_{-0.384}^{+0.556} $ & $0.042_{-0.008}^{+0.003} $ & $0.597_{-0.238}^{+0.319} $ & $0.030_{-0.009}^{+0.013} $ & $0.809_{-0.082}^{+0.072} $ & $0.035\pm 0.006 $ & 0.89 & 3.03\\
\smallskip
38.13& $0.963_{-0.254}^{+0.206} $ & $> 0.034                 $ & $0.950_{-0.751}^{+0.087} $ & $0.049_{-0.039}^{+0.006} $ &--  &  --&  --& --\\
\smallskip
39.03& $9.883_{-3.639}^{+0.117} $ & $0.019\pm 0.005          $ & $1.075_{-0.039}^{+0.038} $ & $0.068\pm 0.001          $ &  --& -- &--  &-- \\
\hline\noalign{\smallskip}
\end{tabular} 
\end{center}
\label{Tab:pds1ni}
Notes:\\
rms: root mean square; $\nu_{\mr{max}}$: characteristic frequency;  $\sigma$: significance; BLN: band limited noise; PN: peaked noise
\end{table*}

\begin{table*}
\caption{Parameters of the QPOs of the PDS}
\begin{center}
\begin{tabular}{rrrrrr}
\hline\noalign{\smallskip}
 \multicolumn{6}{c}{\swift/XRT (0.3 -- 10 keV)}\\ 
\hline\noalign{\smallskip}
 \multicolumn{1}{c}{day} & \multicolumn{1}{c}{$\nu_{\mr{0; QPO1}}$} & \multicolumn{1}{c}{$\Delta_{\mr{QPO1}}$} & \multicolumn{1}{c}{rms$_{\mr{QPO1}}$} & \multicolumn{1}{c}{Q$_{\mr{QPO1}}$}  & \multicolumn{1}{c}{$\sigma_{\mr{QPO1}}$}\\
\hline\noalign{\smallskip}
4.03   & $  0.168_{-0.004}^{+0.006} $&$  0.008_{-0.004}^{+0.005} $&$  0.110_{-0.020}^{+0.018} $&  10.39 &  2.72  \\
\smallskip
6.28   & $  0.449_{-0.011}^{+0.012} $&$  0.131_{-0.033}^{+0.027} $&$  0.095_{-0.013}^{+0.010} $&    1.72  &  3.61 \\
\smallskip
9.27   & $  1.888_{-0.016}^{+0.016} $&$  0.121_{-0.020}^{+0.021} $&$  0.055_{-0.004}^{+0.003} $&  7.80 &  7.10  \\
\smallskip
10.26   & $  2.150_{-0.014}^{+0.013} $&$  0.115_{-0.021}^{+0.027} $&$  0.057\pm0.004 $&  9.36 &  8.10  \\
\smallskip
11.01   & $  2.698_{-0.020}^{+0.019} $&$  0.141_{-0.024}^{+0.033} $&$  0.047\pm0.003 $&  9.55 &  7.83  \\
\smallskip
12.93   & $  2.222\pm0.013 $&$  0.116_{-0.020}^{+0.023} $&$  0.051_{-0.003}^{+0.002} $&  9.59 &  9.14  \\
\smallskip
13.39   & $  2.309_{-0.019}^{+0.018} $&$  0.114_{-0.019}^{+0.021} $&$  0.044\pm0.003 $&  10.13 &  7.32  \\
\smallskip
14.12   & $  2.295\pm0.012 $&$  0.107_{-0.022}^{+0.020} $&$  0.049_{-0.003}^{+0.002} $&  10.70 &  7.19  \\
\smallskip
15.18 & $  2.630_{-0.015}^{+0.016} $&$  0.196\pm0.025 $&$  0.047\pm0.002 $&  6.71 &  10.30\\
\smallskip
16.18  & $  3.220\pm0.019 $&$  0.162_{-0.024}^{+0.026} $&$  0.042\pm0.002 $&  9.94 &  9.43  \\
\smallskip
29.74 & $5.594_{- 0.075}^{+ 0.081} $&$	0.310_{- 0.120}^{+ 0.219} $&$	0.026_{- 0.004}^{+ 0.005}$& 9.04 & 3.06 \\
\smallskip
30.74 & $6.093_{- 0.044}^{+ 0.054} $&$	0.162_{- 0.066}^{+ 0.125} $&$	0.030\pm0.003$& 18.78 & 3.40 \\
\smallskip
32.74  & $  4.360_{-0.031}^{+0.034} $&$  0.201_{-0.041}^{+0.051} $&$  0.046\pm0.003 $&  10.84 &  6.98  \\
\smallskip
35.19  & $  6.475_{-0.070}^{+0.060} $&$  0.165_{-0.080}^{+0.110} $&$  0.024\pm0.004 $& 19.60  &  2.81  \\
\smallskip
37.45  & $  4.503_{-0.039}^{+0.042} $&$  0.170_{-0.054}^{+0.049} $&$  0.034_{-0.005}^{+0.003} $&  13.21 &  3.28  \\
\smallskip
39.32  & $  4.012_{-0.043}^{+0.041} $&$  0.271_{-0.049}^{+0.057} $&$  0.043_\pm0.003 $&  7.42 &  7.38  \\
\smallskip
50.99 & $  3.569_{-0.053}^{+0.052} $&$  0.435_{-0.065}^{+0.080} $&$  0.046\pm0.003 $&  4.10  &  8.23  \\
\smallskip
52.98 & $  2.506_{-0.016}^{+0.032} $&$  0.074_{-0.032}^{+0.053} $&$  0.031_{-0.004}^{+0.005} $&  16.91 &  3.79  \\
\hline\noalign{\smallskip} 
 \multicolumn{1}{c}{day} & \multicolumn{1}{c}{$\nu_{\mr{0; QPO2}}$} & \multicolumn{1}{c}{$\Delta_{\mr{QPO2}}$} & \multicolumn{1}{c}{rms$_{\mr{QPO2}}$}   &  \multicolumn{1}{c}{Q$_{\mr{QPO2}}$}  & \multicolumn{1}{c}{$\sigma_{\mr{QPO2}}$}  \\
\hline\noalign{\smallskip}
9.27   & $  3.756_{-0.038}^{+0.044} $&$  0.162_{-0.056}^{+0.076} $&$  0.034\pm0.005 $&  11.63  & 3.57  \\
\smallskip
10.26  & $  4.416_{-0.095}^{+0.105} $&$  0.392_{-0.144}^{+0.188} $&$  0.038\pm0.006 $&  5.63  & 3.06   \\
\smallskip
11.01   & $  5.576_{-0.170}^{+0.776} $&$  0.474_{-0.189}^{+3.033} $&$  0.030_{-0.006}^{+6.074} $&  5.88  & 2.51  \\
\smallskip
12.93   & $  4.410\pm0.049 $&$  0.200_{-0.064}^{+0.086} $&$  0.028\pm0.004 $&  11.03  & 3.64  \\
\smallskip
13.39   & $  4.738_{-0.085}^{+0.004} $&$  0.001_{-0.001}^{+0.111} $&$  0.013_{-0.003}^{+6.091} $&  21.27  & 2.31  \\
\smallskip
14.12   & $  4.649_{-0.066}^{+0.067} $&$  0.376_{-0.117}^{+0.125} $&$  0.039_{-0.006}^{+0.004} $&  6.18  & 3.37  \\
\smallskip
15.18 & $ 5.340_{-0.085}^{+0.097} $&$  0.519_{-0.163}^{+0.208} $&$  0.031_{-0.005}^{+0.004} $&   5.14  & 3.43  \\
\smallskip
50.99 & $  0.700_{-0.000}^{+0.026} $&$  0.321_{-0.103}^{+0.121} $&$  0.033\pm0.009 $&   2.18  & 1.93  \\
\hline\noalign{\smallskip}
 \multicolumn{1}{c}{day} & \multicolumn{1}{c}{$\nu_{\mr{0; QPO3}}$} & \multicolumn{1}{c}{$\Delta_{\mr{QPO3}}$} & \multicolumn{1}{c}{rms$_{\mr{QPO3}}$}& \multicolumn{1}{c}{Q$_{\mr{QPO3}}$}  & \multicolumn{1}{c}{$\sigma_{\mr{QPO3}}$}\\
\hline\noalign{\smallskip}
9.27   & $  0.29_{-0.02}^{+0.01}  $&$ 0.03_{-0.01}^{+0.02} $&$  0.027\pm0.005 $& 4.69  & 2.63\\
\smallskip
15.18 & $  0.44_{-0.10}^{+0.03}  $&$ 0.07_{-0.04}^{+0.23} $&$  0.020_{-0.005}^{+6.083} $&  2.94  & 2.07\\
\hline\noalign{\smallskip}
 \multicolumn{6}{c}{\xmm\ (1 -- 10 keV)}\\ 
\hline\noalign{\smallskip} 
 \multicolumn{1}{c}{day} & \multicolumn{1}{c}{$\nu_{\mr{0; QPO1}}$} & \multicolumn{1}{c}{$\Delta_{\mr{QPO1}}$} & \multicolumn{1}{c}{rms$_{\mr{QPO1}}$}   &  \multicolumn{1}{c}{Q$_{\mr{QPO1}}$}  & \multicolumn{1}{c}{$\sigma_{\mr{QPO1}}$}  \\
\hline\noalign{\smallskip}
5.81 &  $0.359_{-0.006}^{+0.007}$ & $0.046_{-0.011}^{+0.013}$ &  $0.069\pm0.007$ & 3.94 & 5.21\\
\smallskip
12.73 &$2.345\pm0.007$ & $0.216_{-0.009}^{+0.010}$ & $0.081\pm0.001$ &5.43&31.12\\
\smallskip
14.00 &$2.216_{-0.006}^{+0.007}$ & $0.145_{-0.013}^{+0.014}$ & $0.080\pm0.002$&7.62& 19.10\\
\hline\noalign{\smallskip} 
 \multicolumn{1}{c}{day} & \multicolumn{1}{c}{$\nu_{\mr{0; QPO2}}$} & \multicolumn{1}{c}{$\Delta_{\mr{QPO2}}$} & \multicolumn{1}{c}{rms$_{\mr{QPO2}}$}   &  \multicolumn{1}{c}{Q$_{\mr{QPO2}}$}  & \multicolumn{1}{c}{$\sigma_{\mr{QPO2}}$}  \\
\hline\noalign{\smallskip}
12.73 &$4.516_{-0.032}^{+0.035}$ & $0.326_{-0.326}^{+4.674}$ & $0.044\pm0.002$ &6.93&13.78\\
\smallskip
14.00 &$4.448\pm0.047$ & $0.421_{-0.113}^{+0.130}$ & $0.049\pm0.005$  &5.29&5.19\\
\hline\noalign{\smallskip}
\end{tabular} 
\end{center}
\label{Tab:pds2}
Notes:\\
rms: root mean square; $\nu_0$: centroid frequency; $\Delta$: half width at half maximum; $\sigma$: significance; QPO: quasiperiodic oscillation
\end{table*}

\begin{table*}
\caption{Parameters of the QPOs of the PDS derived from NICER data (0.2 -- 10 keV)}
\begin{center}
\begin{tabular}{rrrrrrc}
\hline\noalign{\smallskip}
 \multicolumn{1}{c}{day} & \multicolumn{1}{c}{$\nu_{\mr{0; QPO1}}$} & \multicolumn{1}{c}{$\Delta_{\mr{QPO1}}$} & \multicolumn{1}{c}{rms$_{\mr{QPO1}}$} & \multicolumn{1}{c}{Q$_{\mr{QPO1}}$}  & \multicolumn{1}{c}{$\sigma_{\mr{QPO1}}$}& \multicolumn{1}{c}{type}\\
\hline\noalign{\smallskip}
10.46   & $ 2.550_{-0.006}^{+0.007}   $&$  0.194_{-0.010}^{+0.011}   $&$  0.066   \pm0.001   $&6.58   &30.09 & C\\
\smallskip
10.99   & $ 2.395\pm0.010   $&$  0.321_{-0.012}^{+0.014}   $&$  0.067   \pm0.001   $&3.73   &33.6 & C  \\
\smallskip
12.00   & $ 1.839\pm0.005   $&$  0.138_{-0.007}^{+0.008}   $&$  0.072   \pm0.001   $&6.66   &29.79 & C \\
\smallskip
13.87   & $ 2.149\pm0.007   $&$  0.105_{-0.008}^{+0.010}   $&$  0.068   \pm0.002   $&10.21   &22.63 & C \\
\smallskip
14.19   & $ 2.595_{-0.009}^{+0.008}   $&$  0.208_{-0.010}^{+0.011}   $&$  0.067   \pm0.001   $&6.23   &30.27 & C \\
\smallskip
15.22   & $ 3.078\pm0.014   $&$  0.308_{-0.010}^{+0.011}   $&$  0.063   \pm0.001   $&5   &39.19 & C \\
\smallskip
16.05   & $ 2.872\pm0.014	                 $&$  0.149\pm0.015                     $&$ 0.065\pm  0.003	 $&	9.64&	12.40 & C\\
\smallskip
16.38   & $ 4.367_{-	0.026}^{+0.030}   $&$  	0.244_{-0.024}^{+0.026}   $&$ 0.053\pm  0.002	 $&	8.95&	13.25 & C\\
\smallskip
16.70   & $ 3.949_{-	0.014}^{+0.016}   $&$  	0.199_{-0.015}^{+0.015}   $&$ 0.057\pm  0.001	 $&	9.90&	22.08 & C\\
\smallskip
16.83   & $ 4.715_{-0.018}^{+0.018}   $&$  	0.274_{-0.025}^{+0.028}   $&$ 0.055\pm  0.002	 $&	8.61&	15.14 & C\\
\smallskip
17.28   & $ 8.992_{-0.065}^{+0.073}   $&$  0.488_{-0.085}^{+0.075}   $&$  0.016   \pm0.001   $&9.22   &7.32 & C \\
\smallskip
18.95   & $ 7.671_{-0.021}^{+0.017}   $&$	0.190_{-0.084}^{+0.072}   $&$	0.021_{-0.002}^{+0.001}   $& 20.15 &6.97 & ?\\
\smallskip
19.01	& $	7.525_{-0.050}^{+0.050}   $&$	0.608_{-0.061}^{+0.054}   $&$	0.028	\pm0.001	$&	6.19	&11.75 & ?\\
\smallskip
19.08	& $	9.335_{-0.077}^{+0.001}   $&$	<0.065   $&$	0.015	\pm0.001	$&	71.37	&12.42 & ?\\
\smallskip
19.15	& $	8.739_{-0.043}^{+0.049}   $&$	0.429_{-0.043}^{+0.043}   $&$	0.021	\pm0.001	$&	10.19	&12.88 & ?\\
\smallskip
19.21	& $	9.335_{-0.001}^{+0.009}   $&$	<0.035   $&$	0.016	\pm0.001	$&	134.52	&16.00 & ?\\
\smallskip
19.34	& $	8.797_{-0.127}^{+0.199}   $&$	0.350_{-0.148}^{+0.138}   $&$	0.015	\pm0.002	$&	12.56	&3.30 & ?\\
\smallskip
19.41	& $	6.831_{-0.068}^{+0.060}   $&$	0.493_{-0.072}^{+0.085}   $&$	0.034	\pm0.002	$&	6.93	&8.92 & ?\\
\smallskip
19.85	& $	5.619_{-0.120}^{+0.129}   $&$	0.263_{-0.215}^{+0.284}   $&$	0.006	\pm0.001	$&	10.69	&2.75 & ?\\
\smallskip
20.37   & $ 6.042\pm0.170   $&$  1.192_{-0.285}^{+0.347}   $&$  0.009   \pm0.001   $&2.53   &5.5 & A? \\
\smallskip
21.01   & $ 5.735_{-0.089}^{+0.088}   $&$  1.137_{-0.118}^{+0.132}   $&$  0.009   \pm0.000   $&2.52   &10.63 & A? \\
\smallskip
22.05   & $ 5.958_{-0.112}^{+0.104}   $&$  0.665_{-0.147}^{+0.201}   $&$  0.006   \pm0.001   $&4.48   &5.33 & A? \\
\smallskip
22.57   & $ 5.053_{-0.053}^{+0.160}   $&$  1.878_{-0.188}^{+0.206}   $&$  0.012   \pm0.001   $&1.35   &7.75 & A? \\
\smallskip
23.02   & $ 5.639_{-0.106}^{+0.105}   $&$  1.196_{-0.155}^{+0.173}   $&$  0.010   \pm0.001   $&2.36   &9.7 & A? \\
\smallskip
24.24   & $ 7.519_{-0.602}^{+0.765}   $&$  2.067_{-0.730}^{+1.008}   $&$  0.006   \pm0.001   $&1.82   &3.81 & A? \\
\smallskip
25.25   & $ 5.295_{-0.147}^{+0.159}   $&$  1.044_{-0.193}^{+0.245}   $&$  0.009   \pm0.002   $&2.54   &7.67 & A? \\
\smallskip
26.74   & $ 5.438_{-0.209}^{+0.208}   $&$  1.464_{-0.269}^{+0.289}   $&$  0.011   \pm0.001   $&1.86   &5.89 & A?\\
\smallskip
28.73   & $ 7.048_{-0.033}^{+0.032}   $&$  0.598_{-0.044}^{+0.046}   $&$  0.025   \pm0.001   $&5.89   &13.72  & C\\
\smallskip
29.76   & $ 5.426\pm0.018   $&$  0.340_{-0.027}^{+0.024}   $&$  0.045   _{-0.002}^{+0.001}   $&7.97   &14.13 & C \\
\smallskip
30.72   & $ 5.726_{-0.012}^{+0.011}   $&$  0.492_{-0.014}^{+0.015}   $&$  0.044   \pm0.001   $&5.82   &44.3 & C \\
\smallskip
31.75   & $ 6.780_{-0.023}^{+0.022}   $&$  0.526_{-0.032}^{+0.033}   $&$  0.033   \pm0.001   $&6.44   &18.5  & C \\
\smallskip
32.71   & $ 4.608_{-0.009}^{+0.010}   $&$  0.355_{-0.017}^{+0.019}   $&$  0.049   \pm0.001   $&6.49   &40.75 & C \\
\smallskip
33.36   & $ 4.897\pm0.014   $&$  0.491_{-0.028}^{+0.033}   $&$  0.044   \pm0.001   $&4.98   &31.07 & C \\
\smallskip
35.03   & $ 5.986_{-0.028}^{+0.025}   $&$  1.177_{-0.047}^{+0.057}   $&$  0.033   \pm0.001   $&2.54   &41.63 & C \\
\smallskip
37.16   & $ 4.706\pm0.021   $&$  0.445_{-0.020}^{+0.020}   $&$  0.040   \pm0.001   $&5.29   &33.08 & C \\
\smallskip
38.13   & $ 4.751\pm   0.019   $&$  0.592_{-0.024}^{+0.027}   $&$  0.049   \pm0.001   $&4.01   &30.75 & C \\
\smallskip
39.03   & $ 4.463\pm0.018   $&$  0.349_{-0.018}^{+0.020}   $&$  0.050   \pm0.001   $&6.39   &25.2 & C \\
\hline\noalign{\smallskip}
\end{tabular} 
\end{center}
\label{Tab:pds2ni}
Notes:\\
rms: root mean square; $\nu_0$: centroid frequency; $\Delta$: half width at half maximum; $\sigma$: significance; QPO: quasiperiodic oscillation
\end{table*}

\addtocounter{table}{-1}
\begin{table*}
\caption{continued}
\begin{center}
\begin{tabular}{rrrrrr}
\hline\noalign{\smallskip} 
 \multicolumn{1}{c}{day} & \multicolumn{1}{c}{$\nu_{\mr{0; QPO2}}$} & \multicolumn{1}{c}{$\Delta_{\mr{QPO2}}$} & \multicolumn{1}{c}{rms$_{\mr{QPO2}}$}   &  \multicolumn{1}{c}{Q$_{\mr{QPO2}}$}  & \multicolumn{1}{c}{$\sigma_{\mr{QPO2}}$}  \\
\hline\noalign{\smallskip}
10.46   & $ 5.084_{-0.024}^{+0.022}   $&$  0.397_{-0.045}^{+0.051}   $&$  0.035\pm0.002   $&6.4   &10.21 \\
\smallskip
10.99   & $ 4.881_{-0.033}^{+0.029}   $&$  1.237_{-0.058}^{+0.061}   $&$  0.050\pm0.001  $ &1.97   &24.85 \\
\smallskip
12.00   & $ 3.616 \pm 0.015 $&$  0.321_{-0.041}^{+0.060}   $&$ 0.041_{-0.003}^{+0.004}   $&5.62	&6.78\\
\smallskip
13.87   & $ 4.302_{-0.023}^{+0.024}   $&$  0.430_{-0.045}^{+0.049}   $&$  0.042\pm0.001  $ &5.01   &16.19 \\
\smallskip
14.19   & $ 5.216_{-0.027}^{+0.026}   $&$  0.887_{-0.056}^{+0.062}   $&$  0.045\pm0.001   $&2.94   &20.59 \\
\smallskip
15.22   & $ 6.162_{-0.030}^{+0.031}   $&$  1.250_{-0.052}^{+0.055}   $&$  0.045\pm0.001   $&2.47   &28 \\
\smallskip
16.05   & $ 5.739_{-0.043}^{+0.045}   $&$  0.710_{-0.093}^{+0.110}    $&$ 0.040 \pm	0.001	$&	4.04	&14.36\\ 
\smallskip
16.38   & $ 9.083_{-0.076}^{+0.080}   $&$  0.837_{-0.118}^{+0.179}    $&$ 0.030 \pm	0.001	$&	5.43	&10.64\\ 
\smallskip
16.70   & $ 7.938_{-0.042}^{+0.043}   $&$  0.688_{-0.065}^{+0.074}    $&$ 0.034 \pm	0.001	$&	5.77	&17.00 \\
\smallskip
16.83   & $ 9.742_{-0.070}^{+0.073}   $&$  0.837_{-0.083}^{+0.086}    $&$ 0.029 \pm	0.001	$&	5.82	&12.95 \\
\smallskip
17.28   & $ 17.733_{-0.768}^{+0.644}   $&$  1.563_{-0.744}^{+0.741}   $&$  0.006   \pm0.001   $&5.67   &3.15 \\
\smallskip
18.95  & $6.463_{-0.104}^{+0.109}   $&$	0.755_{-0.097}^{+0.091}   $&$	0.022\pm	0.002$& 4.28&7.33 \\
\smallskip
19.01	& $ 5.592_{-0.109}^{+0.149}   $&$	0.686_{-0.153}^{+0.193}   $&$	0.016	\pm0.002	$&4.08	&5.03\\
\smallskip
19.08	& $ 7.787_{-0.073}^{+0.079}   $&$	0.730_{-0.112}^{+0.118}   $&$	0.016	\pm0.001	$&5.33	&10.19\\
\smallskip
19.15	& $ 6.650_{-0.142}^{+0.149}   $&$	0.872_{-0.128}^{+0.149}   $&$	0.015	\pm0.001	$&3.81	&6.95\\
\smallskip
19.21	& $ 7.663_{-0.145}^{+0.136}   $&$	0.990_{-0.153}^{+0.160}   $&$	0.014	\pm0.001	$&3.87	&9.00\\
\smallskip
19.34	& $ 6.788_{-0.086}^{+0.107}   $&$	0.542_{-0.264}^{+0.178}   $&$	0.025_{-0.005}^{+0.002}   $&	6.27	&2.57\\
\smallskip
19.34	& $ 5.163_{-0.198}^{+0.439}   $&$	0.384_{-0.378}^{+0.485}   $&$	0.011_{-0.004}^{+0.006}   $&	6.73	&1.53\\
\smallskip
19.41	& $ 5.277_{-0.094}^{+0.167}   $&$	0.316_{-0.316}^{+0.260}   $&$	0.015	\pm0.004	$&	8.35&	1.89\\
\smallskip
22.05   & $ 5.958_{-0.112}^{+0.104}   $&$  0.665_{-0.147}^{+0.201}   $&$  0.006   \pm0.001   $&6.85   &2.22 \\
\smallskip
22.57   & $ 5.053_{-0.053}^{+0.160}   $&$  1.878_{-0.188}^{+0.206}   $&$  0.012   \pm0.001   $&1.38   &1.24 \\
\smallskip
28.73   & $ 13.790_{-0.254}^{+0.257}   $&$  1.677_{-0.287}^{+0.317}   $&$  0.010   \pm0.001  $ &4.11   &7.43 \\
\smallskip
29.76   & $ 10.855\pm0.054   $&$  1.153_{-0.093}^{+0.102}   $&$  0.025   \pm0.001   $&4.71   &17.86 \\
\smallskip
30.72   & $ 11.507\pm0.049   $&$  1.651_{-0.070}^{+0.072}   $&$  0.024   \pm0.000   $&3.49   &39.33 \\
\smallskip
31.75   & $ 13.628_{-0.153}^{+0.151}   $&$  2.137_{-0.191}^{+0.199}   $&$  0.016   \pm0.001   $&3.19   &13.58 \\
\smallskip
32.71   & $ 9.186_{-0.027}^{+0.028}   $&$  0.720_{-0.070}^{+0.078}   $&$  0.025   \pm0.001   $&6.38   &17.86 \\
\smallskip
33.36   & $ 9.916_{-0.086}^{+0.085}   $&$  2.323_{-0.150}^{+0.149}   $&$  0.028   \pm0.001   $&2.13   &15.56 \\
\smallskip
35.03   & $ 12.148_{-0.119}^{+0.126}   $&$  1.291_{-0.227}^{+0.267}   $&$  0.011   \pm0.001  $ &4.71   &7.86 \\
\smallskip
37.16   & $ 9.402_{-0.054}^{+0.055}   $&$  1.118_{-0.065}^{+0.068}   $&$  0.024   \pm0.001   $&4.2   &24.3 \\
\smallskip
38.13   & $ 9.349_{-0.072}^{+0.071}   $&$  1.943_{-0.107}^{+0.160}   $&$  0.032   \pm0.001   $&2.41   &22.57 \\
\smallskip
39.03   & $ 9.009\pm0.051   $&$  0.974_{-0.081}^{+0.094}   $&$  0.032   \pm0.001   $&4.63   &15.75 \\
\hline\noalign{\smallskip}
\end{tabular} 
\end{center}
\label{Tab:pds2nic}
Notes:\\
rms: root mean square; $\nu_0$: centroid frequency; $\Delta$: half width at half maximum; $\sigma$: significance; QPO: quasiperiodic oscillation
\end{table*}

\subsubsection{NICER}
The NICER PDS of the first seven observations analysed in this study show two BLN components and a QPO with upper harmonic  (Fig.~\ref{Fig:pds}). Details on the BLN and the QPOs of all NICER observations can be found in Tables~\ref{Tab:pds1ni} and \ref{Tab:pds2ni}, respectively. The characteristic frequency of the QPO increases from around 2 Hz to about 4 Hz. In the then following observation,  taken on day 17, where the rms drops below 5\%, the characteristic frequency of the QPO increases to $9.01^{+0.07}_{-0.08}$ Hz, comparable to the frequency of the upper harmonic seen in the preceding observation. In addition, the PDS of this observation shows a shoulder to the QPO at lower frequencies ($\nu_{\mr{char}}=6.98^{+0.29}_{-0.28}$ Hz), an upper harmonic at  $\nu_{\mr{char}}=17.80^{+0.83}_{-0.71}$ Hz, and a peaked noise component at $\nu_{\mr{char}}=1.55^{+0.18}_{-0.15}$ Hz. In the next observation, on day 18 the rms drops below 2\% and the PDS is dominated by one BLN component  (Fig.~\ref{Fig:pds}). After an observation gap of 13.7\,h, the rms has increased to 6.1\% and the PDS shows QPOs at $\sim7.7$ and $\sim6.5$ Hz  in addition to the peaked noise component, indicating that \max15\ returned to the hard state. A quite similar PDS is observed at the beginning of the observation taken on day 19. During this observation the characteristic frequencies of the QPOs increase to $\sim9.3$ and $\sim7.8$ Hz and then decrease to $\sim6.8$ and $\sim5.3$ Hz. The characteristic frequency of the peaked noise component increases to $\sim1.5$ Hz and the decreases to $\sim1.1$ Hz. The characteristic frequency of the BLN components are $3.3-6.1$ and $0.9-1.3$ Hz. In the last part of this observation, after a gap of $\sim9$\,h, the rms drops below 2\%, there is one BLN component with a characteristic frequency of $\sim1.3$ Hz and a QPO with a characteristic frequency $\sim5.6$ Hz with a low Q factor of 2.8, indicating that the source made another transition to the soft state. In the then following eight observations the rms is below 2\%, and a QPO with a characteristic frequency in the range of 5 to 8 Hz with rather low Q factor ($\la2$) is observed  (Fig.~\ref{Fig:pds}). The PDS of the observation taken on day 28.7 has an rms of 5\% and shows a QPO at $\nu_{\mr{char}}=7.07\pm0.04$ Hz with a Q factor $>5$ and an upper harmonic. The then following observations all show a QPO with a characteristic frequency between 7 and 4 Hz and an upper harmonic, except for two observations, on days 34 and 36, where the rms drops below 2\% and the PDS can be described by a single BLN component.  

The observations taken between days 10 and 17 clearly show an anti-correlation between QPO frequency and rms variability (Fig.~\ref{Fig:fchar_rms_ni}). The relation between QPO frequency and rms variability for observations taken on days 18 and 19 seems to follow the same anti-correlation, although these observations do not show an upper harmonic and the QPOs in these observations show a different shape with a shoulder towards lower frequencies. Observations taken on after day 28 follow the same anti-correlation. For observations taken between days 20 and 26 we observe a flat correlation between QPO frequency and rms variability. The limited frequency range in which these oscillations appear, as well as the low rms variability and low Q factor suggest that these QPOs are of type-A. As mentioned in Sect.~\ref{Sec:time_sw} the QPOs for which we observe the anti-correlation are most likely of type-C. Based on NICER data, we can conclude that QPOs observed between days 10 and 17, and after day 28 are type-C QPOs.

\subsubsection{\xmm}
The PDS of the three \xmm\ observations can be fitted with two BLN components and a QPO. In the second and third observation an upper harmonic is also present. Details on the BLN and the QPOs can be found in Tables~\ref{Tab:pds1} and \ref{Tab:pds2}, respectively. The centroid frequencies are consistent with the values obtained from \swift/XRT and NICER observations taken close to the \xmm\ ones.

\section[]{Discussion}
\label{Sec:dis}
The evolution of  \max15\  observed during its outburst in fall 2017 is consistent with that usually observed from black hole X-ray binaries. The light curve and diagnostic diagrams show that \max15\ began its outburst in the hard state, increased in luminosity and then evolved towards the soft state. During this evolution strong type-C QPOs are observed. The characteristic frequency of the QPOs increased during the state transition and it was anti-correlated to total fractional rms, which supports the type-C nature of these QPOs. The QPO observed on day 9.27 has already be reported in \citet[][at the same frequency]{2017ATel10734....1M}. The frequency range of the QPOs and their upper harmonics in the then following observations are consistent between \swift/XRT and NICER observations. The QPO frequencies obtained for observations taken on September 12th and 13th are consistent with the values reported in \citet{2017ATel10768....1G}. ALMA and ATCA observations taken on September 11th and 12th, respectively, detected radio emission from a compact synchrotron jet \citep{2017ATel10745....1T}, supporting the finding from the timing studies that \max15\ was in a hard state at the time of these observations. 

The \swift/XRT PDS of observations taken between days 18.97 and 28.75 are dominated by power-law noise and suggest that \maxi15\ has been in the soft state during these observations. In the NICER data we even observe QPOs in observations taken during this part of the outburst. Based on an amount of rms variability below 2\% and the observed flat correlation between QPO characteristic frequency and rms variability, NICER data give additional support that \max15\ has been in the soft state between days 20.36 and 26.73. These findings imply that \max15\ shows an unusually short soft state with a duration of only $\sim$7 days. In the archetypical low-mass black hole X-ray binary \gx339, HSSs with a duration of $\ga100$ days are observed \citep{2002MNRAS.329..588K,2011MNRAS.418.2292M,2011MNRAS.418.1746S}. It is also known that the hard-to-soft state transition is not a smooth transition and often excursions toward higher hardness ratios are observed \citep[\eg][]{2011MNRAS.418.2292M,2013MNRAS.429.2655S}. Interestingly, this softening can not be seen in the hardness-intensity diagram as the hardness ratio does not decrease with increasing intensity. However, it does show up in the rms-intensity diagram.

In the 12 \swift/XRT observations taken between days 29.74 and 50.99 we observed seven QPOs with a characteristic frequency $>3.7$ Hz, which is higher than the characteristic frequencies observed during the brightening of the source. For these seven observations the characteristic frequency of the QPO is above the characteristic frequency of the noise component. In these observations the total fractional rms is 5 -- 9\%, which is in the range where typically type-B QPOs are observed. It is worth noting that the seven \swift/XRT observations seem to correspond to dips in the light curve and to excursions toward higher hardness ratios. Ten NICER observations taken between days 28.7 and 39.0 show QPOs with a characteristic frequency $>4.4$ Hz. Moreover, the characteristic frequency during these observations is higher than that during outburst rise and the characteristic frequency of the QPO is above the characteristic frequency of the noise component. The NICER data also confirm that the total fractional rms is 5 -- 10\%. As can be seen from Fig.~\ref{Fig:pds} the \swift/XRT PDS are much noisier than the NICER PDS and the QPOs are less prominent in the \swift/XRT PDS. For the four NICER PDS where the \swift/XRT PDS taken on the same day do not show a QPO, we fit the \swift/XRT PDS adding a Lorentzian with the centroid frequency set to the value obtained from the NICER PDS, and find that this feature is not significant in the \swift/XRT PDS and that the parameters of all components are not well constrained. For observations of the different instruments taken several hours apart we cannot rule out that \max15\ changed its state between the observations, as the NICER observations show that state changes on time scales of about a day can take place.  The NICER PDS also reveal an upper harmonic for the QPO and they show that the correlation between the characteristic frequency of the QPO and the amount of rms variability follow the same anti-correlation as observed during outburst rise. These findings imply that these QPOs are of type-C and that \maxi15\ has been in a hard state during these observations, despite the total fractional rms value which would be consistent with type-B QPOs. Based on the NICER data we also find that \max15\ shows oscillations during the soft state. These oscillations follow a flat correlation in the rms versus centroid frequency plot (Fig.~\ref{Fig:fchar_rms_ni}), as has been observed for type-A and B QPOs \citep{2011MNRAS.418.2292M}. The oscillations are observed in an rms range (of the total PDS) where type-A QPOs have been observed in studies of RXTE data, but as we observe type-C QPOs at lower rms values than what we expect from RXTE studies, the rms value is not conclusive in determining the QPO type. The low Q factor and low rms value of the oscillations themselves and the frequency range in which these oscillations appear are more consistent with these oscillations being type-A than type-B QPOs. This suggests that we observe type-A and type-C QPOs, but no type-B QPOs during this outburst of \max15. The observations of type-C QPOs in a total fractional rms range where type-B QPOs are expected from studies of RXTE data \citep[\eg\ ][]{2011MNRAS.410..679M}, might be related to the different energy ranges covered by the different satellites. PDS derived from RXTE/PCA comprise the 2 -- 25 keV range, which means they do not include effects of the disk emission present at softer energies, that are covered by all three instruments used in this study, while contributions of the reflection component at energies above 10 keV are covered by RXTE, but missing in the data used here.

The oscillation observed on day 52.98 is clearly a type-C QPO, and coincides with the detection of radio emission around that date \citep{2017ATel10899....1R}, indicating that \max15\ returned to the hard state. MAXI/GSC observations taken after October 25th showed that this return to the hard state was only of temporary nature and that \max15\ reached an even softer state around November 27th \citep{2017ATel11020....1S}. As mentioned above it is not uncommon that the hard-to-soft state transition takes several excursion toward higher hardness ratios and transitions where the source lingered some time irresolutely between the soft and intermediate states before finally reaching the HSS \eg\ in \gx339\ \citep{2011MNRAS.418.2292M} or MAXI\,J1543--564 \citep{2012MNRAS.422..679S}. If the true HSS has only been reached around November 27th, the short duration of the soft state reported in this study is not particularly remarkable and all observations reported here are taken during outburst rise. MAXI/GSC observations showed that \max15\ remained in the soft state until mid April and underwent a hard state transition at the end of April \citep{2018arXiv180400800N,2018ATel11682....1N}. In mid May the source was again observed being in the soft state and another soft-to-hard state transition took place towards the end of May \citep{2018ATel11682....1N}.

The \swift/XRT spectra can be well fitted with an absorbed thermal Comptonization model and do not show indications of a reflection component that has been observed in \nus\ data \citep{2018ApJ...852L..34X}. The non-detection of the reflection component in the \swift/XRT data can be mainly addressed to the different energy ranges covered by the instruments, as the reflection hump shows up between 10 -- 20 keV, a range not covered by \swift/XRT. Furthermore, the rather short exposures of the \swift\ monitoring observations do not have a high enough signal-to-noise ratio to allow us to detect deviation from the Comptonization model at higher energies. We obtain a high foreground absorption of \nh$\sim3$\hcm{22}, which is much higher than what is observed in most black hole X-ray binaries, and about a factor 2 higher than the foreground absorption of \h1743\ \citep[see][for foreground absorptions of several black hole XRBs]{2015MNRAS.452.3666S}. This high foreground absorption hints at \max15\ being an X-ray binary observed at high inclination.


\acknowledgments
This project is supported by the Ministry of Science and Technology of
the Republic of China (Taiwan) through grants 105-2112-M-007-033-MY2 and 105-2811-M-007-065.
This research has made use of data obtained through the High Energy Astrophysics Science Archive Research Center Online Service, provided by the NASA/Goddard Space Flight Center. This work made use of data supplied by the UK Swift Science Data Centre at the University of Leicester, of observations obtained with \xmm, an ESA science mission with instruments and contributions directly funded by ESA Member States and NASA. 
This research has made use of data obtained through the High Energy Astrophysics Science Archive Research Center Online Service, provided by the NASA/Goddard Space Flight Center.

{\it Facilities:} \facility{Neil Gehrels Swift Observatory},  \facility{\xmm},  \facility{NICER}.

\bibliographystyle{apj}
\bibliography{/Users/holger/shao/papers/my2010,/Users/holger/work/papers/my2013}




\end{document}